\documentclass[fleqn,usenatbib]{mnras}
\usepackage{newtxtext,newtxmath}

\usepackage[T1]{fontenc}

\DeclareRobustCommand{\VAN}[3]{#2}
\let\VANthebibliography\thebibliography
\def\thebibliography{\DeclareRobustCommand{\VAN}[3]{##3}\VANthebibliography}

\usepackage{graphicx}	
\usepackage{amsmath}	
\usepackage{verbatim} 

\newcommand{\THI}{\mathrm{\Delta}T_\mathrm{HI}}

\newcommand{\Fmat}{\boldsymbol{F}}
\newcommand{\Mmat}{\boldsymbol{M}}
\newcommand{\Cmat}{\boldsymbol{C}}
\newcommand{\muvec}{\boldsymbol{\mu}}
\newcommand{\xvec}{\boldsymbol{x}}

\title[Weak lensing and intensity mapping cross-correlation]{The feasibility of weak lensing and 21cm intensity mapping cross-correlation measurements}

\author[Sangka, A. \& Bacon, D.]{
Anut Sangka$^{1,2,3}$\thanks{E-mail: anut.sangka@port.ac.uk}
and David Bacon$^{1}$
\\
$^{1}$Institute of Cosmology and Gravitation, University of Portsmouth,  Portsmouth PO1 3FX, UK\\
$^{2}$National Astronomical Research Institute of Thailand, Chiangmai 50180, Thailand,\\
$^{3}$ Department of Physics, Faculty of Science, Ubon Ratchatani University, Ubon Ratchatani 34190, Thailand
}

\date{Accepted XXX. Received YYY; in original form ZZZ}

\pubyear{2023}

\begin{document}
\label{firstpage}
\pagerange{\pageref{firstpage}--\pageref{lastpage}}
\maketitle

\begin{abstract}
One of the most promising probes to complement current standard cosmological surveys is the HI intensity map, i.e. the distribution of temperature fluctuations in neutral hydrogen. In this paper we present calculations of the 2-point function between HI (at redshift $z<1$) and lensing convergence ($\kappa$). We also construct HI intensity maps from N-body simulations, and measure 2-point functions between HI and lensing convergence. HI intensity mapping requires stringent removal of bright foregrounds, including emission from our galaxy. The removal of large-scale radial modes during this HI foreground removal will reduce the HI-lensing cross-power spectrum signal, as radial modes are integrated to find the convergence; here we wish to characterise this reduction in signal. We find that after a simple model of foreground removal, the cross-correlation signal is reduced by $\sim$50-70\%; we present the angular and redshift dependence of the effect, which is a weak function of these variables. We then calculate S/N of $\kappa$HI detection, including cases with cut sky observations, and noise from radio and lensing measurements. We present Fisher forecasts based on the resulting 2-point functions; these forecasts show that by measuring $\kappa\Delta{T}_\mathrm{HI}$ correlation functions in a sufficient number of redshift bins, constraints on cosmology and HI bias will be possible.
\end{abstract}
 


\section{Introduction}

The clustering of matter in the Universe provides an important insight into the origins and evolution of the cosmic structure. Inflation predicts that early structure formation generates a near-Gaussian random field in overdensity; evolution due to gravity causes late-time large-scale structures to exhibit non-Gaussian features. Two point statistics of the density field at different redshifts capture information about the evolution of structures, and correlation functions between different pairs of cosmological probes can precisely constrain cosmological parameters ~\citep{DES:cosmoY1,Troster:2022,DES3Clustering_cosmo,Fang:2022,Upham_2019}.      
Two dimensional surveys of the cosmic microwave background (CMB) have been effectively carried out through the last few decades~\citep{Planck:2018:overview,Hinshaw:2013}. The complement to this is deep sky observations of the 3-dimensional galaxy and dark matter fields. While conventional optical and infrared surveys have high angular resolution, long integration times are needed for these to obtain precise redshifts via spectroscopy. In contrast, photometric surveys provide faster redshift capture but less radial resolution~\citep{Fernandez:2001}. 

To complement the low radial resolution of optical photometric surveys, alternative techniques with higher radial resolution are desirable; radio 21cm intensity mapping is a rapidly developing candidate for this purpose. Unlike most optical surveys, this technique does not measure the brightness of individual objects, but focuses on the larger-scale fluctuations in intensity of the 21cm radio signal from neutral hydrogen (HI). The temperature fluctuations can be used as a tracer for the underlying cosmic density field. This intensity mapping is a complementary technique to a photometric survey, with excellent redshift resolution but lower angular resolution~\citep{Bull:2015}. Hence, combining HI and optical surveys is potentially valuable, as the two techniques compensate for each other's limitations ~\citep{Cunnington:2019a,Bacon:2018dui}. 

Recently, HI intensity mapping techniques have been actively developed~\citep{Santos:2010,Harker:2010,Mao:2008,CHIME:2022a,Wolz:2016,Cunnington:2022a}. The Canadian Hydrogen Intensity Mapping Experiment (CHIME)  ~\citep{CHIME:2022b} has provided a detection of HI via cross-correlations with three probes of Large-Scale Structure (LSS), namely luminous red galaxies (LRG), emission line galaxies (ELG), and quasars (QSO) from the eBOSS clustering catalogs at high significant levels, $7.1\sigma$ (LRG),  $5.7\sigma$ (ELG), and  $11.1\sigma$ (QSO). ~\cite{Cunnington:2022a} have detected the correlated clustering between MeerKAT measurements of HI and galaxies from the WiggleZ Dark Energy Survey at $7.7\sigma$ significance. Intensity mapping is therefore on its way to becoming an independent observational probe, providing useful information from low to high redshifts, via future surveys with radio telescopes such as MeerKAT~\citep{Pourtsidou:2017a,Pourtsidou:2017b,Spinelli:2022} and the Square Kilometre Array, SKA~\citep{Santos:2015}. 

The major challenge for the intensity mapping technique is that the foreground signals are much stronger than the cosmic HI brightness temperature, especially due to the galactic plane synchrotron radiation~\citep{Spinelli_2018,Su_2018,Switzer_2013}. Hence several studies of 2-point functions between HI and optical ($\Delta_\mathrm{HI}\delta_\mathrm{g}$) have focused on the impact of foreground removal~\citep{Chapman:2012,Cunnington:2019a,Padmanabhan_2020,Cunnington_2020,Spinelli:2022}.  The study by ~\cite{Cunnington:2019a} shows that the foreground removal affects 2-point function characteristics, especially when the redshift resolution is broad, as is the case in optical photometric surveys. 

There are also numerous optical surveys measuring gravitational lensing shear ($\gamma$) which distorts the shape of galaxy images; this is sensitive to density fluctuations of all the matter present along a line of sight, whether baryonic or dark matter. It is therefore of interest to consider the viability of the cross-correlation $\gamma-\delta_\mathrm{HI}$, which will be able to be studied using a combination of lensing and IM surveys~\citep{DES:cosmoY1,DES2018_2points_lensing,Hu:2003pt,CHIME:2022a,CHIME:2022b,Cunnington:2022a}. The  density projection along the unperturbed light ray trajectory, also known as 'lensing convergence' $\kappa$ can be considered instead of $\gamma$ as both share the same statistical properties.  The 2-point functions between the pairs of $\kappa$ and HI could improve cosmological constraints and break degeneracies such as that between HI bias ($b_\mathrm{HI}$) and clustering amplitude.

However, removing the HI foreground potentially affects these 2-point statistics, as the foreground removal effectively subtracts large-scale radial modes to which lensing is sensitive.  In this paper we will calculate the cross-correlation function between convergence and 21cm intensity mapping, and will explore whether the foreground subtraction significantly hampers the cross-correlation measurement. We also explore whether the foreground removal impacts the viability of cosmological constraints from HI-HI and $\kappa-$HI correlations. 

To achieve this, we will present theoretical and simulation approaches for calculating the $\kappa$-HI signal. We will then consider the effect of foreground removal on the signal, showing that the impact is significant (approximately a factor of 2 in signal reduction) but not lethal. We will then use the Fisher information matrix to make cosmological parameter forecasts for ideal and realistic surveys (including cut sky and the inclusion of telescope-specific noise), deploying the cross-correlation between convergence and intensity mapping, always including the effect of foreground removal.
We discuss lensing convergence and HI simulation catalogues in Section~\ref{sec:sim}, including modeling of the  2-point functions. We describe the HI foreground removal and its effect on $\kappa$-HI 2-point functions in Section~\ref{sec:foreground}. We present our Fisher forecasts for surveys in Section~\ref{sec:fisher}, effects of instrumental noise in  Section~\ref{sec:instrument_noise} and present our conclusions in  Section~\ref{sec:conclusions}.

\section{$\kappa$-HI 2-point statistics: theory and simulations}\label{sec:sim}
In this section we discuss the relevant 2-point statistics. We shall start with theoretical calculations of 2-point functions of lensing convergence ($\kappa$) and neutral hydrogen intensity maps (HI) in subsection~\ref{subsec:cl_model}. We will then discuss the generation of $\kappa$ catalogues and HI modelling from simulations of the matter overdensity $\delta$. The comparison between theoretical calculations and simulations is shown in subsection~\ref{subsec:maps}. The simulated HI maps will be used in the next section~\ref{sec:foreground} where the foreground removal will be discussed.
\subsection{Modeling the 2-point functions}\label{subsec:cl_model}
In this subsection, we describe the modeling of the 2-point functions. We begin by considering how to calculate the observable quantities, namely weak lensing convergence $\kappa$ and HI temperature fluctuations $\THI$. We will then turn to the angular cross-power spectra. We denote $\kappa\kappa$ as the power spectra between $\kappa$ fields, HI$^{i}$HI$^{j}$ as the cross-power spectra between HI fields, and $\kappa$HI as the cross-power between $\kappa$ and HI. The dummy indices $i$ and $j$ refer to the $i$th and $j$th redshift bins. We will calculate the lensing convergence in an arbitrary direction on the sky $\hat{n}$ using the Born approximation, projecting the matter overdensity $\delta$ along an unperturbed ray direction. This can be computed by~\citep{Bertelmann:2001}
\begin{equation}\label{eq:kappa}
    \kappa(\chi_s,\hat{n}) = \frac{3\Omega_\mathrm{m}H^2_0}{2c^2} \int_0^{\chi_s} d\chi' \frac{\chi'(\chi-\chi)}{\chi}\frac{\delta(\hat{n},\chi')}{a(\chi')},
\end{equation}
where $\chi$ is comoving distance, $\Omega_{\rm m}$ is the matter density parameter at the present epoch, $H_0$ is the Hubble parameter today and the subscript $s$ refers to the source plane. For lensing of distributed sources in redshift bins $i$, the integrand is modified by including a source distribution, so that the integration now becomes
\begin{equation}\label{eq:kappa_i}
    \kappa^i(\hat{n}) = \int_0^\infty d\chi' q^{i}_\kappa (\chi')\delta(\hat{n},\chi'),
\end{equation}
where the lensing weight is given by
\begin{equation}\label{eq:kappa_weight}
    q^{i}_\kappa(\chi) = \frac{3\Omega_\mathrm{m}H^2_0}{2c^2} \int_0^{\chi_s} \frac{\delta(\hat{n},\chi')}{a(\chi')} \int_\chi^\infty d\chi' \frac{\chi-\chi}{\chi}\frac{n^i_s(z(\chi'))\frac{dz}{d\chi'}}{\bar{n}^i_s},
\end{equation}
where $n^i_s(z)$ is the lensing source number density, and $\bar{n}^i_s$ is its average in the $i$th redshift bin.

HI will be a biased tracer of matter overdensity, so we write $\THI$ $(\hat{n},z)$ = $\bar{T}_\mathrm{HI}(z)b_\mathrm{HI}(z)\delta(\hat{n},z)$, where $b_\mathrm{HI}(z)$ is the HI bias at a given redshift $z$ and $\bar{T}_\mathrm{HI}(z)$ is the average temperature. The projected temperature fluctuation at the $i$th redshift bin is then 
\begin{equation}\label{eq:deltaTi}
\THI^{i}(\hat{n}) = \int_0^{\chi_i} d\chi' q^i_\mathrm{HI} (\chi')\delta(\hat{n},\chi'),
\end{equation}
where 
\begin{equation}\label{eq:HI_weight}
    q^i_\mathrm{HI}(\chi) = \bar{T}_\mathrm{HI}(z(\chi)) b^i_\mathrm{HI}(\chi) \frac{n^i_\mathrm{HI}(z(\chi))\frac{dz}{d\chi}}{\bar{n}^i_\mathrm{HI}},
\end{equation}
where $n^i_\mathrm{HI}(z)$ is the HI source number density, and $\bar{n}^i_\mathrm{HI}$ is its average in the $i$th redshift bin. 

\citet{Battye:2013} show that for a given redshift $z$, $\bar{T}_\mathrm{HI}(z)$ can be estimated by
\begin{equation}\label{eq:T_bar}
    \bar{T}_\mathrm{HI}(z) = 44 \mu\mathrm{K} \bigg(\frac{\mathrm{\Omega}_\mathrm{HI}(z)h}{2.45 \times 10^{-4}}\bigg) \frac{(1+z)^2}{E(z)},
\end{equation}
where $E(z) = H(z)/H_0$ is the dimensionless Hubble function at redshift $z$. The HI density parameter could be approximated to be $\mathrm{\Omega}_\mathrm{HI}h = 2.45\times10^{-4}$ ~\citep{Battye:2013}. However, throughout this research we shall follow the fitting formula for the SKA-MID I by~\cite{Bacon:2018dui}
\begin{equation}
\mathrm{\Omega}_\mathrm{HI}(z) = 0.00048 + 0.00039z - 0.000065z^2.
\end{equation}
Constraining the HI bias $b_\mathrm{HI}(z)$ will be discussed later in section~\ref{sec:fisher}.

Using the Limber approximation, the angular power spectra $C^{XY}(\ell)$ are given by
\begin{equation}\label{eq:limber}
  C^{XY}(\ell) = \int d\chi \frac{ q_X(\chi)q_Y(\chi)}{\chi^2}P_\delta \bigg ( \frac{\ell+ 1/2}{\chi},z(\chi)   \bigg),
\end{equation}
where $P_\delta \bigg ( \frac{\ell+ 1/2}{\chi},z(\chi)   \bigg)$ is the matter power spectrum~\citep{LoVerde:2008}. We compute the nonlinear power spectrum using the Boltzmann code CAMB~\citep{Lewis:2002ah} with the Halofit extension to nonlinear scales~\citep{Takahashi:2012}.

Since the ray tracing simulations by~\cite{Takahashi:2017a} which we use below adopt a comoving bin size $\rm\Delta{\chi}$ = 150 $h^{-1}$Mpc (see section~\ref{subsec:maps}), we  choose a radial selection function for $n^i_\mathrm{HI}(z(\chi))/\bar{n}^i_\mathrm{HI}$ as a normal distribution around a central comoving position with $3\sigma_\chi$ = 150 $h^{-1}$Mpc.  This $\sigma_\chi$ corresponds to the frequency bandwidth $(\Delta\nu)$ selected. In practice, the frequency range and bandwidth will depend on the particular radio telescope being used; for example, BINGO (Baryon Acoustic Oscillations in Neutral Gas Observations) has operational frequency from 960 MHz to 1260 MHz~\citep{Battye:2013,BINGO:2019} and MeerKAT's frequency bandwidth ranges from 900-1185 MHz and 580-1000 MHz for L-band and UHF-band, respectively~\citep{Wang:2021,Cunnington:2022a} . 

\begin{figure}
  \centering
   {
   \includegraphics[width=0.5\textwidth]{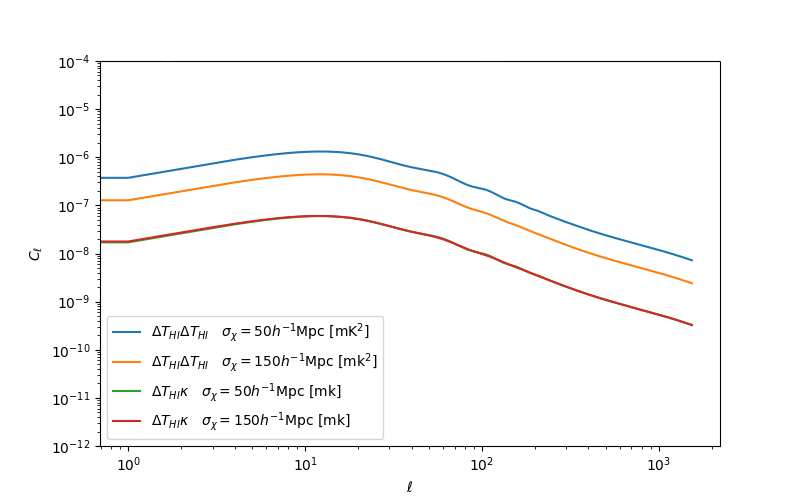}
    \caption{Power spectra $C_\ell$ for HI and cross-power between HI and convergence, for radial HI width $\sigma_\chi$ = 50 and 150 $h^{-1}$Mpc.  The effect of the width is less important for the cross-correlation. Different $\sigma_\chi$ corresponds to different frequency bandwidths, $\Delta{\nu}$ of the radio data.}\label{fig:binning}
    }
\end{figure}

We show examples for the first time of calculations of the auto- and cross-power for HI and convergence in Fig.~\ref{fig:binning} using Eq.~\ref{eq:limber}. As expected, the auto- signal depends on the radial HI width $\sigma_\chi$, while the cross-power is insensitive to this.

\subsection{Lensing Convergence and HI Intensity Maps}\label{subsec:maps}
The full-sky gravitational lensing mock catalogues by ~\cite{Takahashi:2017a} have been used throughout this work.  They are based on a multiple-lens ray-tracing approach through N-body cosmological simulations. The datasets include weak lensing maps (convergence, shear and rotation data) up to redshift 5.3, and halo catalogues. The catalogues provide 108 realisations of N-body simulations, 35 of which are used in this research (due to storage limitations). The N-body simulations were produced with periodic boundary conditions following dark matter gravitational evolution without baryonic processes. 14 simulation boxes of side length $L$ = 450, 900, 1350, ..., 6300 $h^{-1}$Mpc are nested to represent a region of the Universe in which lensing occurs; each box contains $2048^3$ particles. The $\kappa$ fields are obtained by tracing the light ray path through planes with separation 150 $h^{-1}$Mpc. By calculating the Jacobian matrix $A$ along the light path, the lensing convergence $\kappa$, shear lensing $\gamma_{1,2}$ and rotation angle $\omega$ can be obtained, via
\begin{gather}
 A
 =
  \begin{bmatrix}
  1-\kappa -\gamma_1 &
   -\gamma_2 - \omega \\
   -\gamma_2 + \omega  &
   1-\kappa + \gamma_1 
   \end{bmatrix}.
\end{gather}
 The convergence maps were created in the HEALPIX scheme with NSIDE of 4096~\citep{Healpix:2005}, which contain 200 megapixels. While this resolution is appropriate to study nonlinear structure and matches forthcoming galaxy surveys such as EUCLID~\footnote{https://www.euclid-ec.org/} and DESI~\footnote{https://www.desi.lbl.gov/}, the cross-correlation between the lensing convergence and the HI intensity map is limited by the lower angular resolution of HI intensity maps expected with real radio telescopes. Therefore the resolution is reduced to NSIDE of 512; this is not only appropriate for our 2-point function measurements but also decreases the storage space requirement and computational time. 
 
We will first consider a convergence map at a specific optical lensing catalogue source redshift, which we choose as $z\approx$ 0.78, for which the lensing will significantly occur at the redshift of an intensity map at redshift$\simeq 0.3$. This particular choice of redshift allows us to compare our results to current and forthcoming optical and radial surveys~\citep{DES2018_2points_lensing,Bacon:2018dui,Euclid:2019VII,Pourtsidou:2017b,Santos:2015}. We will then extend to multiple lensing planes (see Table \ref{tab:redshift_bin}).

We turn now to generating our IM maps. Crucially, we will emulate removal of the IM foreground by removing the radial temperature fluctuations on large scales. The foreground removal will be discussed in detail in Section~\ref{sec:foreground}. 

\begin{figure}
  \centering
        \includegraphics[width=0.3\textwidth]{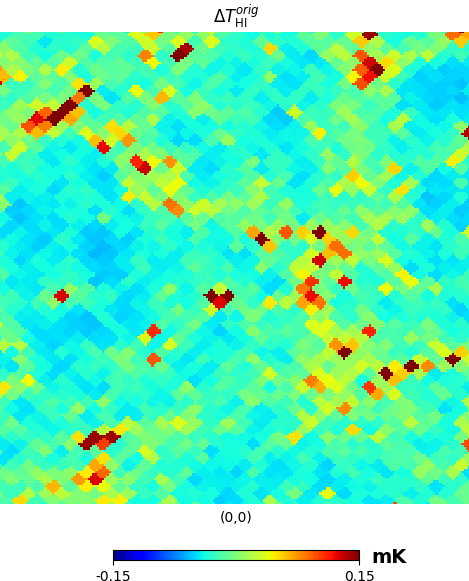}
        \includegraphics[width=0.3\textwidth]{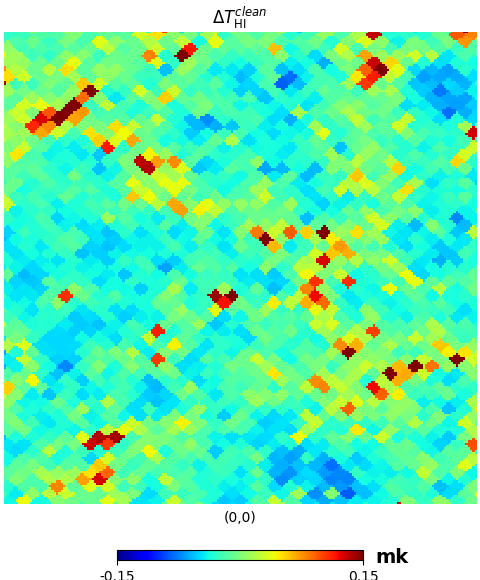}
        \includegraphics[width=0.3\textwidth]{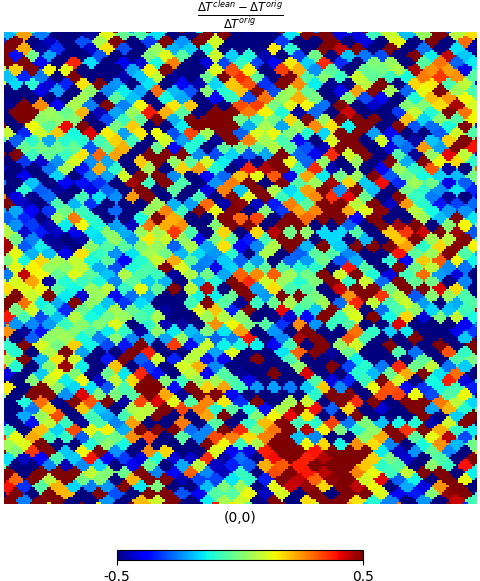}
    \caption{Top: the uncleaned intensity map, $\Delta{T}_\mathrm{HI}^\mathrm{orig}$, at $z=0.3$ from an example realisation. The fluctuations were measured by assuming $b_\mathrm{HI}(z)$ (see Eq.~\ref{eq:HI_fluc} and~\ref{eq:HI_bias}). 
 Middle: the foreground-removed intensity map, $\Delta{T}_\mathrm{HI}^\mathrm{clean}$, at the same redshift. The foregrounds were removed by eliminating radial long wavelength modes up to redshift $z_\mathrm{max}=1$. The NSIDEs of the fluctuation maps is reduced from 4096 to 512 to match the resolution of forthcoming radio surveys. Bottom: residual map of cleaned and uncleaned maps. Each of these detail maps has area 5 $\times$ 5 square degrees (a small patch of the entire sky maps).
}
\label{fig:maps}
\end{figure}

First we need to make the pre-foreground-removal IM maps. Instead of calculating the individual HI masses $M_\mathrm{HI}$ from halo catalogues, we assume that HI is a biased tracer of the total matter overdensity field $\delta(\theta,z)$(see Eq.~\ref{eq:deltaTi} and~\ref{eq:HI_weight}), 
\begin{equation}\label{eq:HI_density}
\delta_\mathrm{HI}(\hat{n},z) \equiv \frac{T_\mathrm{HI}(\hat{n},z)-\bar{T}_\mathrm{HI}(z)}{\bar{T}_\mathrm{HI}(z)} = b_\mathrm{HI}(z)\delta(\hat{n},z),
\end{equation}
where $b_\mathrm{HI}$ is a HI bias. For instance the parametric form for $b_\mathrm{HI}$ adopted by ~\cite{Cunnington:2019a} is
\begin{equation}\label{eq:steve_HI}
b_\mathrm{HI}(z) = 0.67+0.18z+0.05z^2 .
\end{equation}
Since the neutral hydrogen signal is measured as the surface brightness temperature, we shall refer to the HI intensity map as the temperature fluctuation $\THI$ :
\begin{equation}\label{eq:HI_fluc}
\THI(\hat{n},z) = T_\mathrm{HI}(\hat{n},z) - \bar{T}_\mathrm{HI}(z) = \bar{T}_\mathrm{HI}(z)b_\mathrm{HI}(z)\delta(\hat{n},z).
\end{equation}
We apply this equation to the overdensity map obtained from~\cite{Takahashi:2017a} catalogues to create HI intensity maps. Fig.~\ref{fig:maps} shows the uncleaned and cleaned intensity maps from one realisation for the zoom-in patch with area 5 $\times$ 5 square degrees. 

As we are interested in the 2-dimensional projection of cosmological fields on the sky, together with their power spectra, it is  convenient to describe these fields $\Theta(\hat{n},z)$ in spherical harmonics:
\begin{equation}\label{eq:Ylm}
\Theta(\hat{n},z) = \sum_{\ell=0}^\infty\sum_{m=-\ell}^{m=\ell}a_{\ell m}(z)Y^{m}_\ell(\hat{n}),
\end{equation}
where $Y^m_\ell(\hat{n})$ and $a_{\ell m}(z)$ are spherical harmonics and their coefficients respectively~\citep{Pratten:2016,Castro:2005,Heavens:2003}. $\Theta(\hat{n},z)$ represents an arbitrary cosmological field; in this work it can be either lensing convergence or HI temperature fluctuations. The angular power spectrum is then an average of $a_{\ell m}$ over $m$ modes:
\begin{equation}\label{eq:C_XY}
C^{XY}(\ell) = \langle a^X_{\ell m} (z_1) a^{Y*}_{\ell m}(z_2) \rangle,
\end{equation}
where $X$ and $Y$ stand for the cosmological fields at given redshifts $z_1$ and $z_2$, respectively. 

The cross power-spectrum for HI and lensing $\kappa$ can be easily measured via HEALPIX's {\tt anafast} routine especially if the data is for the full sky (however, if the data has missing regions or a cut sky, pseudo-$C_\ell$ methods are required~\citep{Brown:2005,Upham_2019}).  

\begin{figure}
  \centering 
     \includegraphics[width=0.50\textwidth]{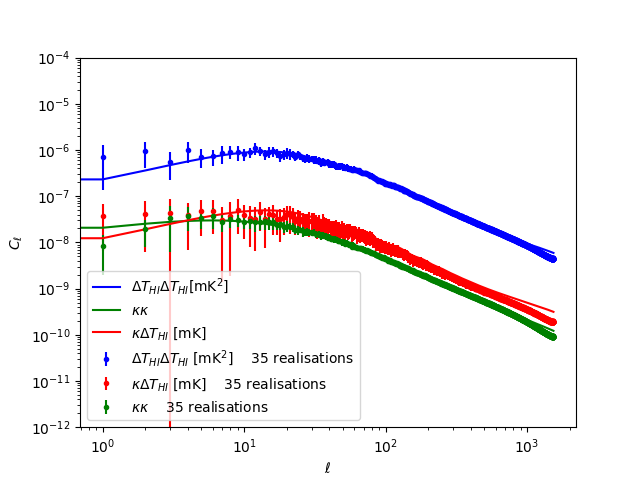}     
  \caption{Comparison between theoretical $C_\ell$ (see Eq.~\ref{eq:limber}) and measured $C_\ell$ from our simulations. Here the lensing convergence is measured at central redshift 0.78 and HI is measured at central redshift 0.3. }\label{fig:lens_shell}
\end{figure}

Using this routine, we obtain cross-power measurements for the HI and $\kappa$ fields. We measure the cross-power spectra of 35 realisations and evaluate their mean; we show the results in Fig.~\ref{fig:lens_shell}. Here the lensing convergence is measured at the central redshift 0.78 and HI is measured at the central redshift 0.3. Fig.~\ref{fig:lens_shell} also displays a comparison between theoretical 2-point statistics and the measurements from the mock catalogues. We then measure the covariance matrices $\mathbf{COV}$($C^{XY}$) of 2-point statistics from 35 realisations. The error bars are the square root of the diagonal elements of  $\mathbf{COV}$($C^{XY}$) of the estimators. The correlation matrix for $\mathbf{COV}$($C^{XY}$) are shown in Fig.~\ref{fig:corr}.  

We see that the measurements from simulations agree very well with our theory curves on this plot, which indicates that our theoretical calculation and selection function $n^i_\mathrm{HI}(z(\chi))/\bar{n}^i_\mathrm{HI}$ successfully match the simulations. Due to the lens shell approximation of the ray tracing code, the measured $C_\ell$ is slightly affected at very high $\ell$ (see red line on Fig.~\ref{fig:lens_shell}).

\section{HI foreground removal and its effect on $\kappa$HI 2-point functions}\label{sec:foreground}
The HI signal is small compared to its foregrounds such as free-free thermal emission, extragalactic radio sources and Galactic synchrotron.  For example, the synchrotron ($T_\mathrm{sync}$) emission temperature which can be modelled by $T_\mathrm{sysnc} \propto (1+z)^{2.7} [\mathrm{K}]$ ~\citep{Platania:1998,Smoot:2017}, is approximately 3 to 4 orders of magnitude larger than $T_\mathrm{HI}$ at low redshift. Thus, 21cm foreground removal is a major challenge for HI cosmology. Several studies suggest that the foreground spectrum appears to be smooth in the radial direction~\citep{Cunnington:2019b,Cunnington:2019a,Shaw:2014}. This is equivalent to being present in the long radial wavelengths in Fourier space. We therefore remove such modes in the line of sight background temperature fluctuations $\THI^\mathrm{LoS}(\hat{n})$. 

Since the calculation of lensing involves integration along the light path (Eq.~\ref{eq:kappa_weight}), which will have a contribution from long-wavelength radial modes, the HI foreground removal is a concern for the existence of the $\kappa$HI cross-correlation (i.e. we have just removed such modes from the HI signal). In this section we therefore seek to ascertain the degree to which the $\kappa$HI cross-correlation survives foreground removal.

Here we follow the method for foreground removal emulation by~\citet{Cunnington:2019a}. The cleaned intensity map $\THI^\mathrm{clean}$  can be approximated as
\begin{equation}\label{eq:clean}
\THI^\mathrm{clean}(\hat{n},z) = \THI^\mathrm{orig}(\hat{n},z)-\THI^\mathrm{LoS}(\hat{n}),
\end{equation}
where $\THI^\mathrm{orig}(\hat{n},z)$ is the uncleaned signal in direction $\hat{n}$ at redshift $z$. $\THI^\mathrm{LoS}(\hat{n})$ is defined by:
\begin{equation}\label{eq:T_LOS}
\THI^\mathrm{LoS}(\hat{n}) =\frac{1}{N_z}\sum_i\bar{T}_\mathrm{HI}(z_i)b_\mathrm{HI}(z_i)\delta(\hat{n},z_i),
\end{equation}
 so that $\THI^\mathrm{LoS}(\hat{n})$ is the mean surface brightness temperature fluctuation along the entire line of sight.  This is an initial very approximate model of Principal Component Analysis (PCA) foreground removal, as most dominant components are included in the line of sight expectation temperature fluctuations  $\THI^\mathrm{LoS}(\hat{n})$. It is worth mentioning that this blind foreground removal technique assumes the smoothness of the foreground. However, this smoothness can be hampered by non-smooth features of the beam, e.g. beamwidth of the radio dish, and some oscillating features in all bands of MeerKAT. A simple $1/f$ dependence of the beam could generate artificial HI signals. This leads to the conclusion in  ~\cite{Spinelli:2022} that it is fundamental to develop accurate beam deconvolution algorithms and test data post-processing steps carefully before cleaning. This topic of beam deconvolution is beyond the scope of our research; here we shall assume that the $1/f$ behaviour is sufficiently small. For more sophisticated foreground cleaning methods  we encourage the reader to explore e.g. ~\cite{Cunnington:2023}.

In this work we adopt the same bias model as~\citet{Cunnington:2019b}:
\begin{equation}\label{eq:HI_bias}
    b_\mathrm{HI}(z) = \alpha(b_0 + b_1 z + b_2 z^2),
\end{equation}
where $\alpha$, $b_0$, $b_1$ and $b_2$ are set to 1, 0.67, 0.18 and 0.05, respectively. ~\citet{Cunnington:2019b} obtained this parameter set by investigating HI as a biased tracer of the large-scale structure via HI intensity map and optical galaxy number density cross-correlations (see Eq. 39 from ~\citet{Cunnington:2019b}). We use this as a fiducial model since the HI redshift range in our work is similar to~\citet{Cunnington:2019a,Cunnington:2019b}. Note that in this model, we solely account for the redshift evolution of HI bias and assume any transverse scale dependence of the bias is negligible.~\cite{Martin:2012} shows that this is a good approximation for scales > 10 $h^{-1}$Mpc, which are our main interest.

We measure $\THI^\mathrm{LOS}$ with two choices of maximum redshift, $z_\mathrm{max}$ = 1 and 3. $z_\mathrm{max}$ = 3 corresponds to futuristic HI-galaxy surveys~\citep{Bacon:2018dui}. On the other hand $z_\mathrm{max}$ = 1 is an approximate limit for HI maps with SKA1-MID and MeerKAT~\citep{Bacon:2018dui,Cunnington:2022a}.

\begin{figure}
  \centering
        \includegraphics[width=0.45\textwidth]{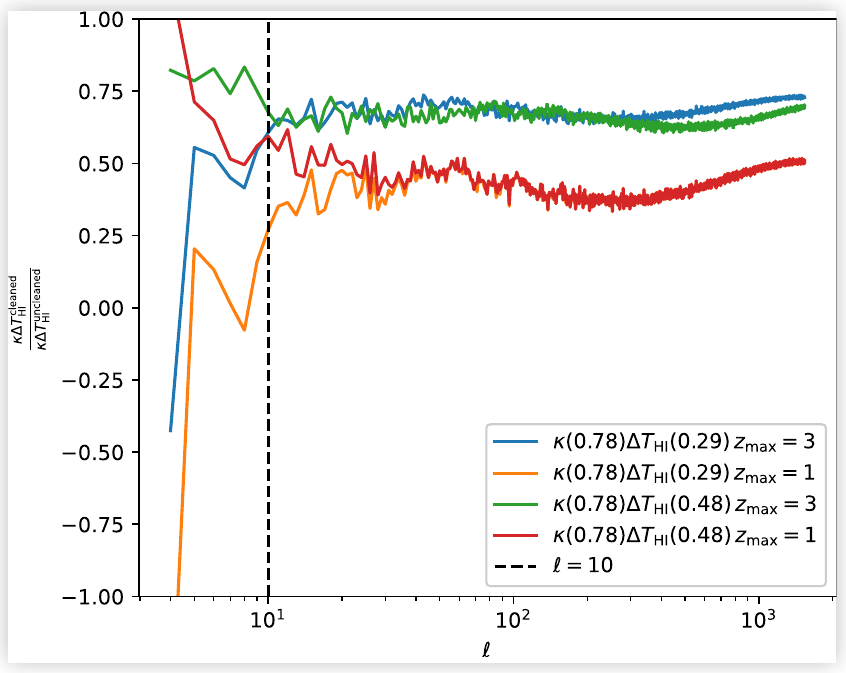}
    \caption{ The ratio between cleaned and uncleaned $\kappa\Delta{T}_\mathrm{HI}$ power spectra. Two maximum redshifts ($z_\mathrm{max})$ for foreground removal are considered; $z_\mathrm{max} =1$  corresponds to current and imminent radio dishes meanwhile $z_\mathrm{max} = 3$ represents a future SKA survey.  }\label{fig:deltacl}
\end{figure}

\begin{figure}
  \centering

        \includegraphics[width=0.5\textwidth]{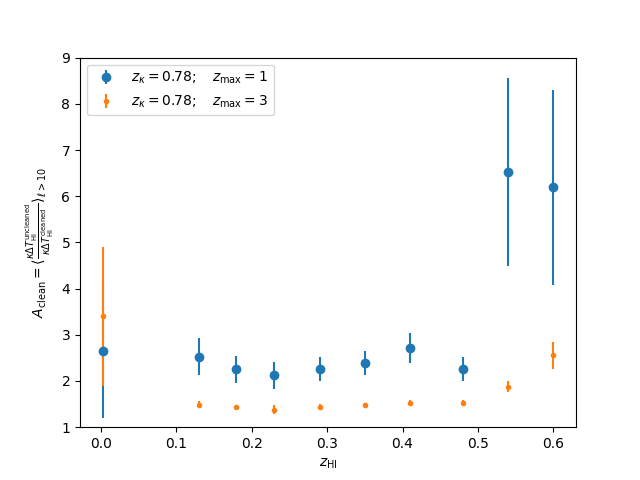}
    \caption{The average ratio $A_\mathrm{clean}$ over $\ell > 10$ modes, as a function of redshift of HI slice used in the cross-correlation.}
        
\label{fig:Aclean}
\end{figure}

We use these intensity maps with removed foreground to calculate the auto-power spectra of the intensity map ($\Delta{T}_\mathrm{HI}\Delta{T}_\mathrm{HI}$), and the cross-power spectra between HI and $\kappa$ ($\kappa\Delta{T}_\mathrm{HI}$). We compare the signal of removed and unremoved  $\kappa\Delta{T}_\mathrm{HI}$, resulting  in Fig.~\ref{fig:deltacl} and Fig. ~\ref{fig:Aclean}. From Fig.~\ref{fig:deltacl}, we note that foreground removal strongly affects the signal on large scales. However we find that on smaller scales,  at $\ell > 10$, the foreground removal does not erase the $\kappa\Delta{T}_\mathrm{HI}$ power spectrum; the signal is scaled down by a factor ($A_\mathrm{clean}$) which is close to constant over a range of $\ell$ modes from 10 to 1000. Hence, in section~\ref{sec:fisher}, when cosmological constraints following foreground removal are considered, the estimation of cosmological parameters is based on the signal where $\ell>10$ . We describe the mean signal drop $A_\mathrm{clean}$ by:
\begin{equation}\label{eq:Aclean}
A_\mathrm{clean}(z_\mathrm{HI}, z_\kappa, z_\mathrm{max}) \equiv \bigg\langle \frac{\kappa\Delta{T}_\mathrm{HI}^\mathrm{uncleaned}}{\kappa \Delta{T}_\mathrm{HI}^\mathrm{cleaned}} \bigg\rangle _{ 10 < \ell < 1500}.
\end{equation}
From Fig.~\ref{fig:deltacl}, we see that the higher the maximum redshift of the survey in which we remove the LOS signal, the less is the effect on the cleaned cross-correlation signal, as more radial modes are preserved in the removal process. Fig.~\ref{fig:deltacl} and~\ref{fig:Aclean} further indicate that the signal in the $\kappa$HI 2-point correlations drops by approximately the same factor $A_\mathrm{clean}$ across a wide redshift range, if we remove the background noise up to a particular redshift $z_\mathrm{max}$, when cross-correlating to the $\kappa$ field at a fixed redshift. Fig.~\ref{fig:Aclean} also implies that $A_\mathrm{clean}(z_\mathrm{HI}, z_\kappa, z_\mathrm{max1}) < A_\mathrm{clean}(z_\mathrm{HI},z_\kappa,z_\mathrm{max2})$ if the maximum redshifts $z_\mathrm{max1}$ > $z_\mathrm{max2}$.

\section{Fisher forecast}\label{sec:fisher}
In previous sections, we have presented the theoretical 2-point statistics for the HI-lensing cross-correlation, and have examined the impact of HI foreground removal on the cross-power spectrum. The results indicate that foreground removal reduces the 2-point statistics by a modest factor.

Here we begin the exploration of $\kappa$HI correlations as a tool for cosmological constraints. In particular we will make a Fisher information matrix forecast of this correlation in the case of low instrument noise  (but including our foreground subtraction model); this will assess the best-case capacity of this probe to constrain cosmology, when one is dominated by large-scale structure fluctuations in the HI and lensing fields.  We will then examine more realistic cases with cut sky and the inclusion of instrumental noise.

\subsection{The Fisher matrix}\label{subsec:F_matrix}
The Fisher information matrix is a useful tool to estimate the expected uncertainty in cosmological parameters for forthcoming experiments~\citep{Heavens:2003,Tegmark:1997}. Assuming that the model parameters $\theta_i$ are distributed by a multivariate Gaussian likelihood $L$, the Fisher matrix can be calculated as
\begin{equation}\label{eq:f_matrix}
    \boldsymbol{F}_{ij} \equiv \bigg < \frac{\partial^2 \mathcal{L}}{\partial \theta_i \partial \theta_j} \bigg >,
\end{equation}
where $\mathcal{L}=-\ln{L}$. The Fisher matrix can be used to obtain the minimum uncertainty ($\sigma_i$) in parameter estimation due to the Cram\'{e}r-Rao inequality~\citep{Mendez:2014,Kamionkowski:2011},
\begin{equation}\label{eq:cramer}
    \sigma_i \geqslant \sqrt{\boldsymbol{F}^{-1}_{ii}},
\end{equation}
which is equivalent to a 68$\%$ confidence level. For a dataset where the uncertainties are Gaussian, the Fisher matrix can be calculated by~\citep{Tegmark:1997}
\begin{equation}\label{eq:f_gaussian}
\boldsymbol{F}_{ij} = \frac{1}{2}\mathrm{Tr}[\boldsymbol{A}_i\boldsymbol{A}_j + \Cmat^{-1}\Mmat_{ij}]
\end{equation}
where $\Cmat$ denotes the covariance matrix of the data, 
$\boldsymbol{A}_i$ $\equiv$ $\Cmat^{-1}\Cmat_{,i}$, 
the derivative data matrix $\Mmat_{ij}$ $\equiv$ $\muvec_{,i}\muvec^{T}_{,j}$ + $\muvec_{,j}\muvec^T_{,i}$, and $\muvec$ is an expectation value of the data vector $\xvec$. The comma symbol means the partial derivative operator with respect to the parameter, $\muvec_{,i}$ $\equiv$ $\partial\muvec/\partial\theta_i$. Note that all derivatives are performed at the maximum likelihood point.

\begin{figure*}
  \centering
        \includegraphics[width=0.32\textwidth]{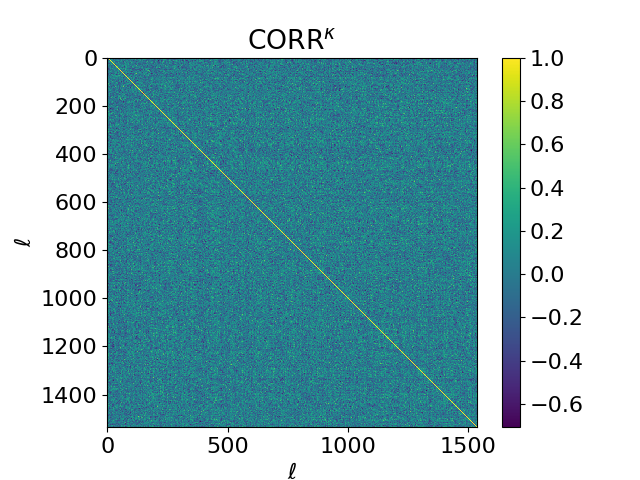}
        \includegraphics[width=0.32\textwidth]{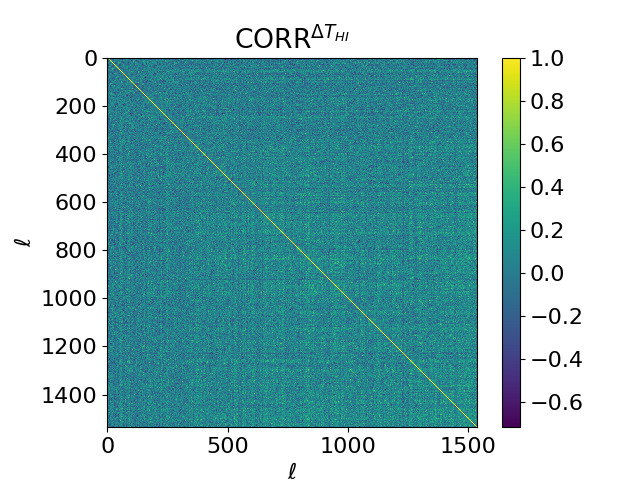}
        \includegraphics[width=0.32\textwidth]{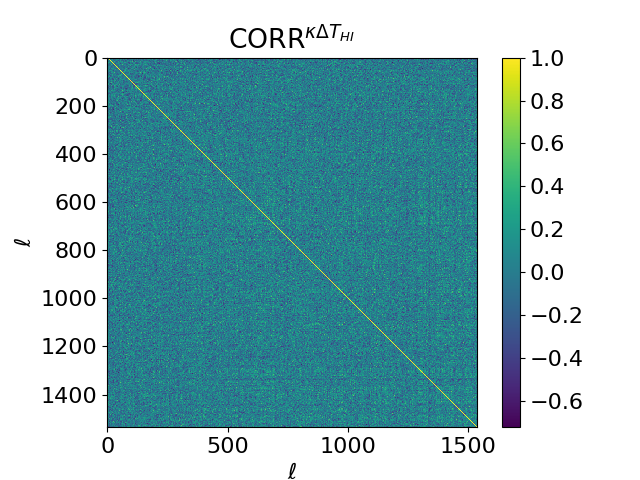}
    \caption{The correlation matrices of $C_\ell^{XY}$ measured from 35 realisations for the cleaned HI signal at central redshift $z=0.3$ and $\kappa$ signal at $z=0.78$. Left: $\kappa\kappa$, middle: $\THI\THI$ and right: $\kappa\THI$. 
    } \label{fig:corr}
\end{figure*}

As we expect only small changes in the covariance matrix $\mathbf{COV}(C^{XY})$ under a modest change in cosmological parameters  (see Sec.~\ref{sec:sim} and~\ref{sec:foreground}), the first term on the right hand side in Eq.~\ref{eq:f_gaussian} will be negligible. Then the Fisher matrix can be written
\begin{equation}\label{eq:f_data}
\Fmat_{ij}=\sum_{XY}\frac{\partial C^{XY}}{\partial \theta_i}^T\mathbf{COV}(C^{XY})^{-1}\frac{\partial C^{XY}}{\partial \theta_j}.
\end{equation}
We calculate the cross-power spectra $C^{XY}(\ell)$ by using Eq.~\ref{eq:limber} with Planck 2018 cosmological parameters~\citep{Planck:2018:overview}. We calculate the covariance matrices of $\kappa\kappa$, $\kappa$HI and HIHI from measured cross power-spectra of 35 realisations of the N-body simulation by~\citet{Takahashi:2017a}. All the HI temperature fluctuation maps which we use take into account foreground removal. We also calculate the correlation matrices $\mathbf{CORR}_{ij}=\mathbf{COV}_{ij}/\sqrt{\mathbf{COV}_{ii}\mathbf{COV}_{jj}}$ and show these in Fig.~\ref{fig:corr}; these do not indicate significant correlations between $\ell$ bins. 
\subsection{Cosmological Constraints for Single-Slice Cross-correlations}\label{subsec:cosmo_1}
In this section, the cosmological constraint viability of $\kappa$ and HI cross-correlations is explored. We first start with the simplest observational configuration, considering only one redshift slice of HI and $\kappa$. We further assume that the $b_\mathrm{HI}(z)$ behaves as in Eq.~\ref{eq:HI_bias}.
We use the Planck 2018 cosmological parameters as the fiducial cosmology~\citep{Planck:2018:overview}. The fiducial cosmological parameters are $h$ = 0.67, $\Omega_\mathrm{m}$ = 0.3, $\sigma_8$ = 0.82, $\Omega_\mathrm{k}$ = 0, $\Omega_\Lambda$ = 0.7, $\tau$  = 0.06, and $n_s$ = 0.96. To make a covariance matrix of cross-power spectra for the Fisher matrix (Eq.~\ref{eq:f_data}), we combine $\ell$ modes into 15 bins; each bin contains 101 $\ell$ modes with $11 \le \ell \ge 1527$ and averages over 35 realisations. We first consider the 3$\times$2 functions for a joint analysis of $\kappa (0.78)\kappa (0.78)$,
$\THI(0.3)\THI(0.3)$ and $\kappa (0.78)\THI (0.3)$, where the  numbers in brackets are the central redshifts. We choose these central redshifts as examples of current HI and lensing surveys' central redshifts. The `2$\times$2' functions refer to the same combination but exclude the weak lensing-HI cross power spectrum. 

\begin{figure*}
\centering
\includegraphics[width=0.45\textwidth]{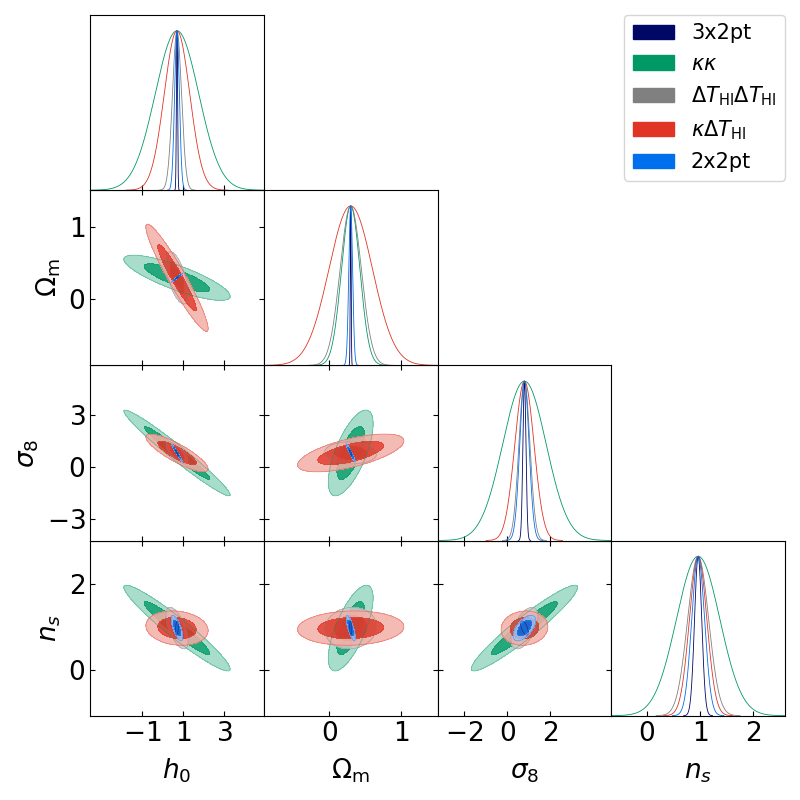}
\includegraphics[width=0.45\textwidth]{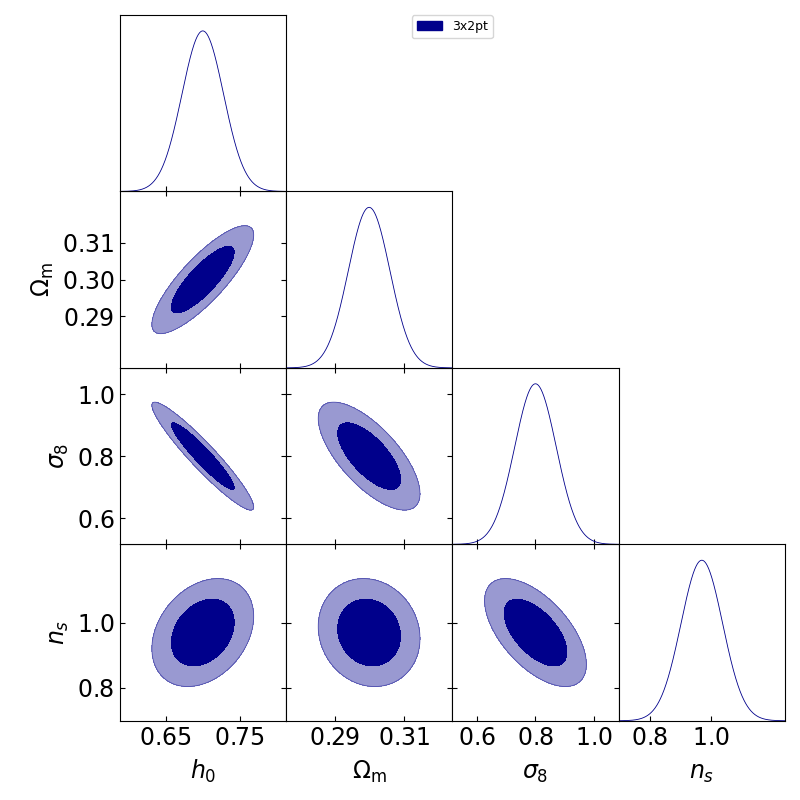}
\caption{Left: Likelihood contours for the dataset described in section~\ref{subsec:cosmo_1};  contours show 68$\%$ and 95$\%$ confidence levels. Right: zoom-in of likelihood contours of 3$\times$2-point functions and their marginalisations from the left panel. }\label{fig:3_set}
\end{figure*}

Fig.~\ref{fig:3_set} shows the joint likelihood obtained via Fisher matrices (see Eq.~\ref{eq:f_data}). We see that single redshift slice correlations of $\kappa-\kappa$(green) and HI-HI(grey) provide relatively weak constraints, while 2x2pt and particularly 3x2pt are more promising, with few- to ten-per cent constraints available on parameters in this low noise case. The zoom-in version of $3 \times 2$ pt functions is shown on the right hand side of Fig.~\ref{fig:3_set}; these likelihood contours, which include the cross-correlation, show a significant improvement in cosmological constraints compared to $\kappa\kappa$ or $\THI$$\THI$ constraints alone. Therefore in the next section, we will examine a joint likelihood between more redshift bins, and where the HI bias ($b_\mathrm{HI}(z)$) is taken into account.

\subsection{HI bias and multi-redshift bin joint likelihood analysis}\label{subsec:bias_analysis}

It is well known that there is a degeneracy between galaxy bias, $\Omega_\mathrm{m}$ and $\sigma_8$ in parameter constraints, since these parameters all affect the amplitude of the power spectrum (see Eq.~\ref{eq:limber}). However, they contribute differently to the  evolution of the power spectrum with time; hence by measuring the power spectra in various redshifts we can break the degeneracies between them. From Eq.~\ref{eq:limber}, we can see that while $\THI\THI$  measures $b_\mathrm{HI}^2(z)$, $\kappa\THI$ additionally measures $b_\mathrm{HI}(z)$. Combining the cross-bin intensity mapping power spectra with the $\kappa\THI$ cross-spectra we can therefore tighten our constraints on bias and cosmological parameters.

In this section we consider two different bias models, with distinct parameter sets. The first, more restricted model explores bias amplitude variation via the $\alpha$ parameter in Eq.~\ref{eq:HI_bias}, setting the rest of the parameters to the best fit values~\citep{Cunnington:2019a}. In the second model, $b_0$, $b_1$ and $b_2$ are the bias parameters with $\alpha$ set to equal 1. We include these  parameters when evaluating the Fisher matrices (Eq.~\ref{eq:f_data}). Note that both $b_\mathrm{HI}$ models are scale-invariant and depend only on $z$.  We will consider both full-sky and 300 deg$^2$ surveys to explore the viability of HIHI and $\kappa$HI in cosmological constraints.

For the full-sky case, as we include more parameters for $b_\mathrm{HI}$, we also examine more redshift bins for both HI and $\kappa$ to obtain the best possible results. We consider the redshift range for $\THI$ which would be measured by pre-SKA and SKA-MID experiments~\citep{Bacon:2018dui,Pourtsidou:2017b,Santos:2015}.   Table~\ref{tab:redshift_bin} shows the central redshifts we consider for $\THI$ and $\kappa$ bins; the width of each bin is 150 $h^{-1}$Mpc, $\Delta{z}$ $\approx$ 0.05. Table~\ref{tab:redshift_bin} lists both HI and $\kappa$ central redshifts. As we have 16 $z_\mathrm{HI}$ , we shall refer to '16-HIHI' which corresponds to 16 pairs of HI auto-correlation functions; we cross-correlate HI intensity maps to $\kappa$ fields at $z_\kappa$ = 0.44, 0.78 and 1.77 respectively.  We refer to 16-HIHI+1-$\kappa$ as the joint analysis for 16-HIHI and $\kappa$HI 2-point statistics at which $z_\kappa$ = 0.44.  We add further $z_\kappa$ bins and label joint data as 16-HIHI+2-$\kappa$ and 16-HIHI+3-$\kappa$.  We calculate both the futuristic case where HI can be measured with high $\ell_\mathrm{max} \geq 1000$ and the current state of art where $100<\ell_\mathrm{max}<400$.

\begin{figure}
    \centering
    \includegraphics[width=0.5\textwidth]{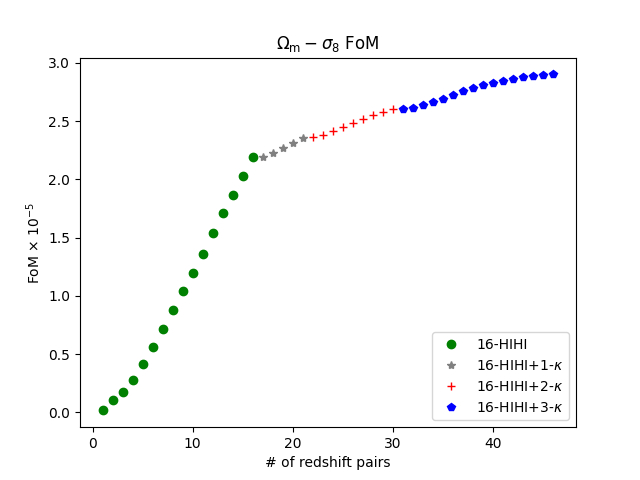}
    \caption{Figure of Merit for $\sigma_8 - \Omega_m$ constraints; the  horizontal axis is the number of redshift bin pairs for cosmological constraints. We show cumulative FoM when including increasing numbers of HI auto-correlation redshift bins (green); then increasing numbers of cross-correlations with convergence bins (grey, red, blue). Here $\ell_\mathrm{max} =1530$.}
    \label{fig:FoM}
    \includegraphics[width=0.5\textwidth]{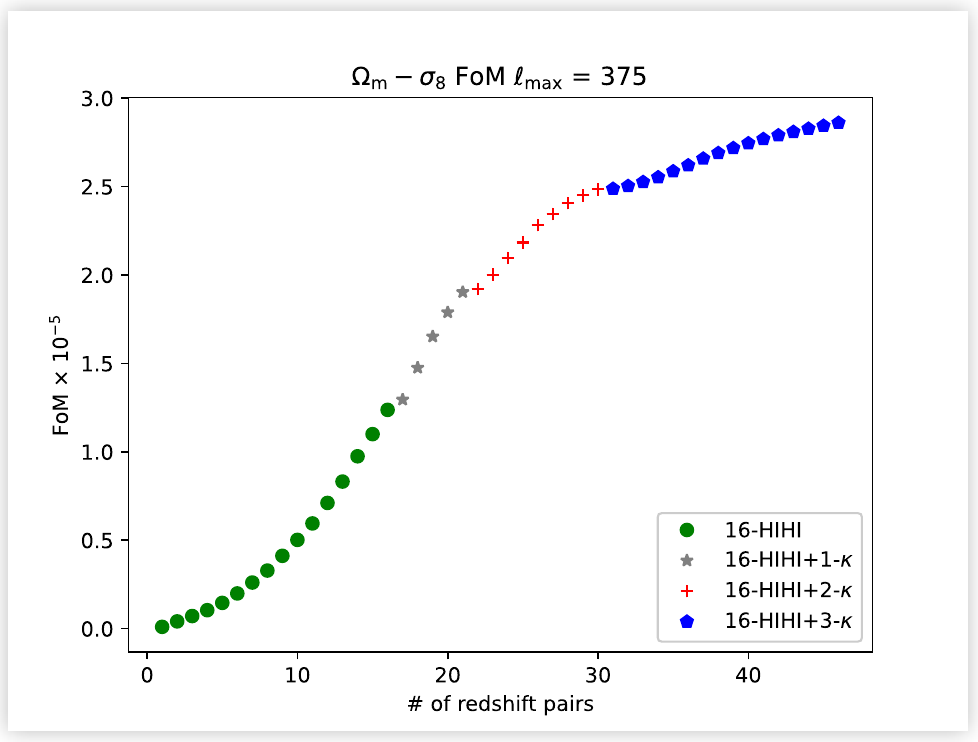}
    \caption{Figure of Merit for $\sigma_8 - \Omega_m$ constraints; the  horizontal axis is the number of redshift bin pairs for cosmological constraints. We show cumulative FoM when including increasing numbers of HI auto-correlation redshift bins (green); then increasing numbers of cross-correlations with convergence bins (grey, red, blue). Here $\ell_\mathrm{max} = 375$.}
    \label{fig:FoMlmax300}
\end{figure}

We further calculate the figure of merit (FoM) for the $\Omega_\mathrm{m}-\sigma_8$ constraint. The FoM is the inverse of the area of the $\Omega_\mathrm{m}-\sigma_8$ contours; in this case we calculate the FoM at 95$\%$ confidence level. Fig.~\ref{fig:FoM} shows the FoM of the $\Omega_\mathrm{m} - \sigma_8$ constraints. The blue dots show the FoM from 16HIHI, where we consecutively add HI auto-correlations for the redshift bins in the order listed in Table~\ref{tab:redshift_bin}. We see that all redshift bins contribute to an improved signal, with a nearly linearly increasing contribution (for this experiment, we are assuming that only $z<1$ HI slices are available). The green, red and black dots in Fig.~\ref{fig:FoM} show the FoM for $\ell_\mathrm{max} = 1530$ when we further add the cross-correlations between consecutive HI slices and the $\kappa$ slices at $z$ = 0.44, 0.78 and 1.78, respectively. We see that these cross-correlations significantly improve the FoM, and appear to be converging to a maximal constraint when including all slices.

\begin{table}
	\centering
	\caption{The central redshifts for intensity and lensing convergence maps in our multi-redshift bin analysis. For $\kappa\THI$, we require that $z_\mathrm{HI} < 0.7z_\kappa$.}
	\label{tab:redshift_bin}
	\begin{tabular}{lccr} 
		\hline
		$z_\mathrm{HI}$ & $z_\kappa$ \\
		\hline
		0.02 & 0.44 \\
		0.08 & 0.78\\
		0.13 & 1.77\\
		0.18 \\
		0.24 \\
		0.29 \\
        0.35 \\
        0.41 \\
        0.47 \\
        0.54 \\
        0.60\\
        0.68\\
        0.75\\
        0.83\\
        0.91\\
        0.99\\
		\hline
	\end{tabular}
\end{table}

\begin{figure}
\centering
\includegraphics[width=0.5\textwidth]{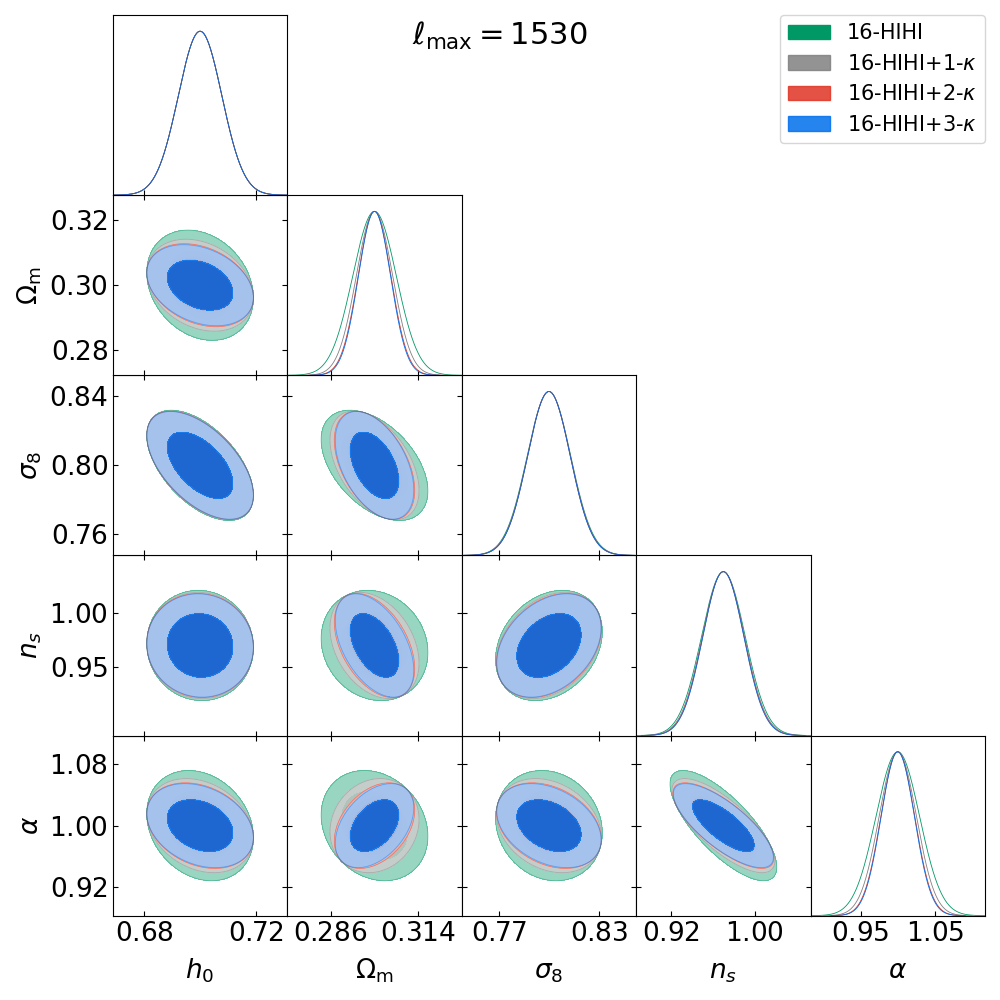}
\caption{The likelihood contours for our multi-bin analysis with HI bias model 1 with $\ell_\mathrm{max} = 1530$; the contours show 68$\%$ and 95$\%$ confidence levels. The cosmological parameter uncertainties at 95$\%$ are measured and reported in Table~\ref{tab:cosmo_param}. }\label{fig:bias_1}
\includegraphics[width=0.5\textwidth]{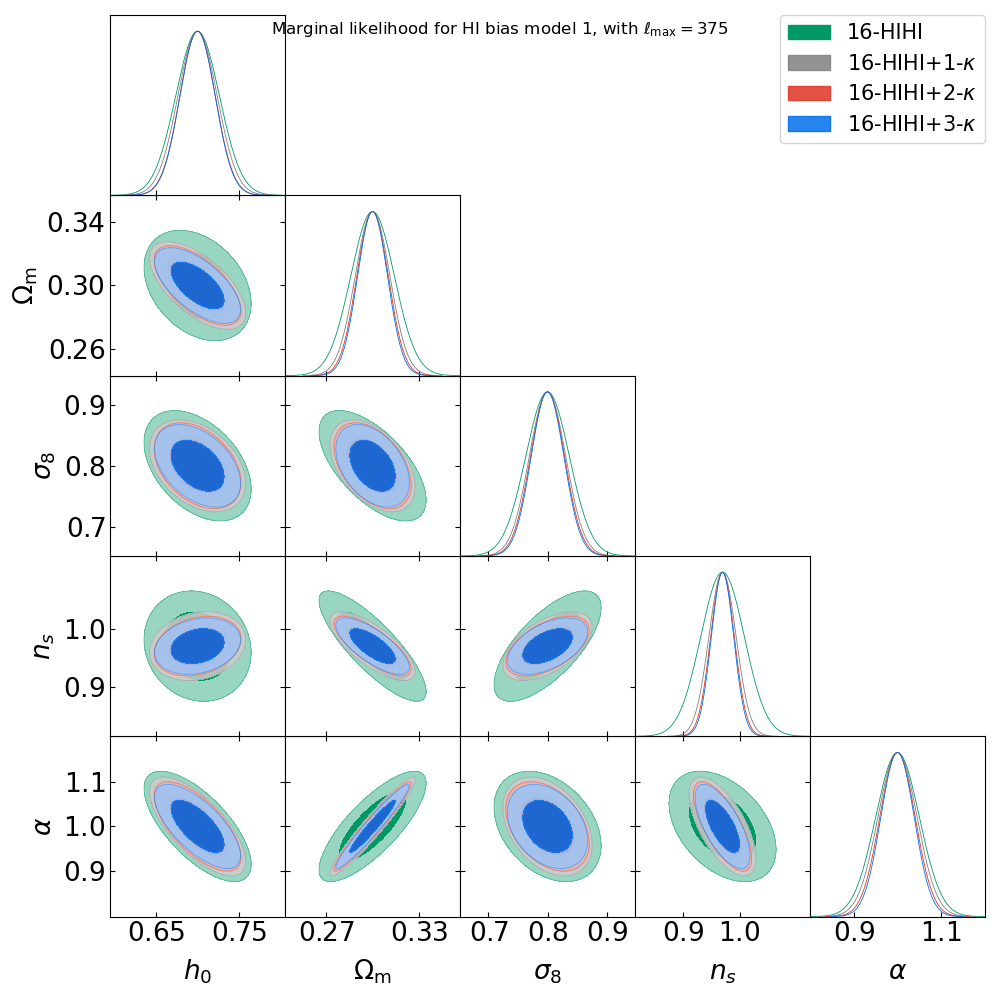}
\caption{The likelihood contours for our multi-bin analysis with HI bias model 1 with $\ell_\mathrm{max}=375$; the contours show 68$\%$ and 95$\%$ confidence levels. Compared to Fig.~\ref{fig:bias_1}, we can see the difference in 16-HIHI constraint. However, when combining the $\kappa\mathrm{HI}$-correlation the constraints are similar to those in Fig.~\ref{fig:bias_1}.  }\label{fig:bias_1l300}
\end{figure}

We now present Fisher forecast results for the first bias model, using the redshift bins in Table~\ref{tab:redshift_bin}. Fig.~\ref{fig:bias_1} shows the utility of multi-redshift bin power spectra measurements with HI and $\kappa$. With the appropriate redshift bin size of HI ($\Delta{z}_\mathrm{HI}$) for $\ell_\mathrm{max} = 1530$, we can achieve tight cosmological constraints (in this low noise case) which are comparable to the optical and CMB probes such as~\citep{Abbott:2018wog,BOSS:cosmo2017,Planck:2018:overview,DES:cosmoY1}. By including more $\kappa$ redshift slices, the
constraints are improved significantly especially for  $\Omega_\mathrm{m}$ and $\alpha$. However,
this also makes the contours more elliptical, as there are remaining degeneracies
among parameters. The uncertainties on parameters are measured at  95$\%$ confidence level and reported in Table~\ref{tab:cosmo_param}.

We next consider the current state of the art case, where since the typical angular resolution $\sim1^\circ$, we set $\ell_\mathrm{max}$ = 375 (we choose this particular value as it is convenient to consider the $\Delta{\ell}$ bins as 15 bins with $\Delta{\ell}$ = 25). Fig.~\ref{fig:FoMlmax300} shows the cumulative Figure of Merit in this case, while Fig.~\ref{fig:bias_1l300} illustrates the likelihood contours for cosmological parameters with this maximum multipole. Comparing with Fig.~\ref{fig:bias_1}, where $\ell_\mathrm{max} = 1530$ we can see that there is a substantial difference in the 16-HIHI contours. However, we notice the significant improvement in parameter constraints when we add 3 $\kappa$ bins to 16-HIHI (green shade) in $\ell_\mathrm{max}$ = 375.  We are therefore seeing that by joining $\kappa$HI 2-point statistics to HIHI auto-correlations, we can improve cosmological constraints significantly.

\begin{table}
	\centering
	\caption{Cosmological forecast from Fisher analysis for 16-HIHI + 3-$\kappa$ correlations; the uncertainties on cosmological parameters and HI bias are quoted at 95$\%$ confidence level. HI bias model 1 considers only the re-scaling parameter $\alpha$. In contrast the second model considers quadratic parameters; $b_2$ is poorly constrained, while all other parameters are able to be measured well (see Fig.~\ref{fig:bias_2bot}).  }
	\label{tab:cosmo_param}
	\begin{tabular}{lccr} 
		\hline
		Parameters & HI bias model 1 & HI bias model 2\\
		\hline
		$\Delta{h_0}$  & $\pm$ 0.02 & $\pm$ 0.02 \\
		$\Delta{\Omega_\mathrm{m}}$ & $\pm$ 0.01& $\pm$ 0.02 \\
		$\Delta{\sigma_8}$ & $\pm$ 0.03 & $\pm$ 0.04\\
		$\Delta{n_s}$ & $\pm$ 0.04 & $\pm$ 0.05 \\
        $\Delta\alpha$ & $\pm$ 0.04 & - \\
        $\Delta{b_0}$ & - & $\pm$ 0.04 \\
        $\Delta{b_1}$ & - & $\pm$ 0.03 \\
        \hline
	\end{tabular}
\end{table}

\begin{figure}
\centering
\includegraphics[width=0.50\textwidth]{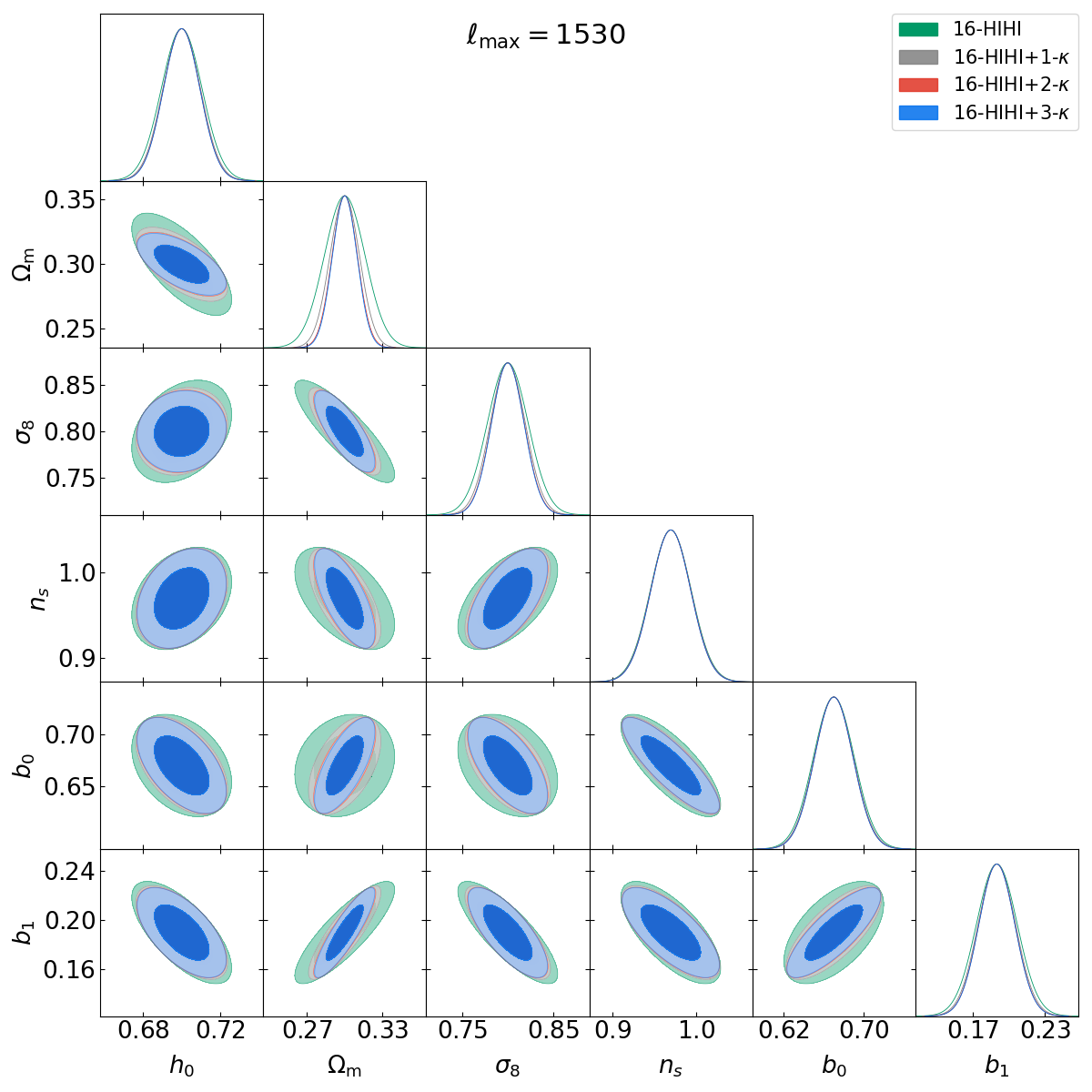}
\caption{Constraints on cosmological parameters and $b_\mathrm{HI}(z)$ for our second bias model, for 2-point functions at 68\% and 95\% levels of confidence (for our 16HIHI redshift slice case). The $\kappa$HI correlation functions do not significantly improve the $h_0$, $n_s$, $b_0$ and $b_1$ constraints. For this figure, we marginalised over $b_2$ as it is poorly constrained. However, we see significant improvement in $\Omega_\mathrm{m}$ and $\sigma_8$ constraints. The parameter uncertainties at 95$\%$ level are reported in Table~\ref{tab:cosmo_param}.}\label{fig:bias_2bot}
\end{figure}

For bias model 2, we find that the second order coefficient of HI bias $b_2$ is very poorly constrained. Marginalising over this parameter does not significantly affect the other cosmological parameter constraints ($h_0$, $\Omega_\mathrm{m}$, $\sigma_8$ and $n_s$). We set this parameter to 0.05 following~\cite{Cunnington:2019a}. Fig.~\ref{fig:bias_2bot} shows the likelihood constraints for this model. We can see that by adding more $\kappa$ slices, the $h_0$ constraints do not improve much but the improvement in the $\Omega_\mathrm{m}-\sigma_8$ constraint can be easily noticed. The uncertainties on our HI bias models and cosmological parameters are reported in Table~\ref{tab:cosmo_param}. Comparing the parameter constraints from both $b_\mathrm{HI}$ models (see Table~\ref{tab:cosmo_param}), we notice that model 1 gives  slightly better (but very comparable) cosmological constraints. 

\subsection{The effect of sky coverage}
\begin{figure}
\centering
\includegraphics[width=0.50\textwidth]{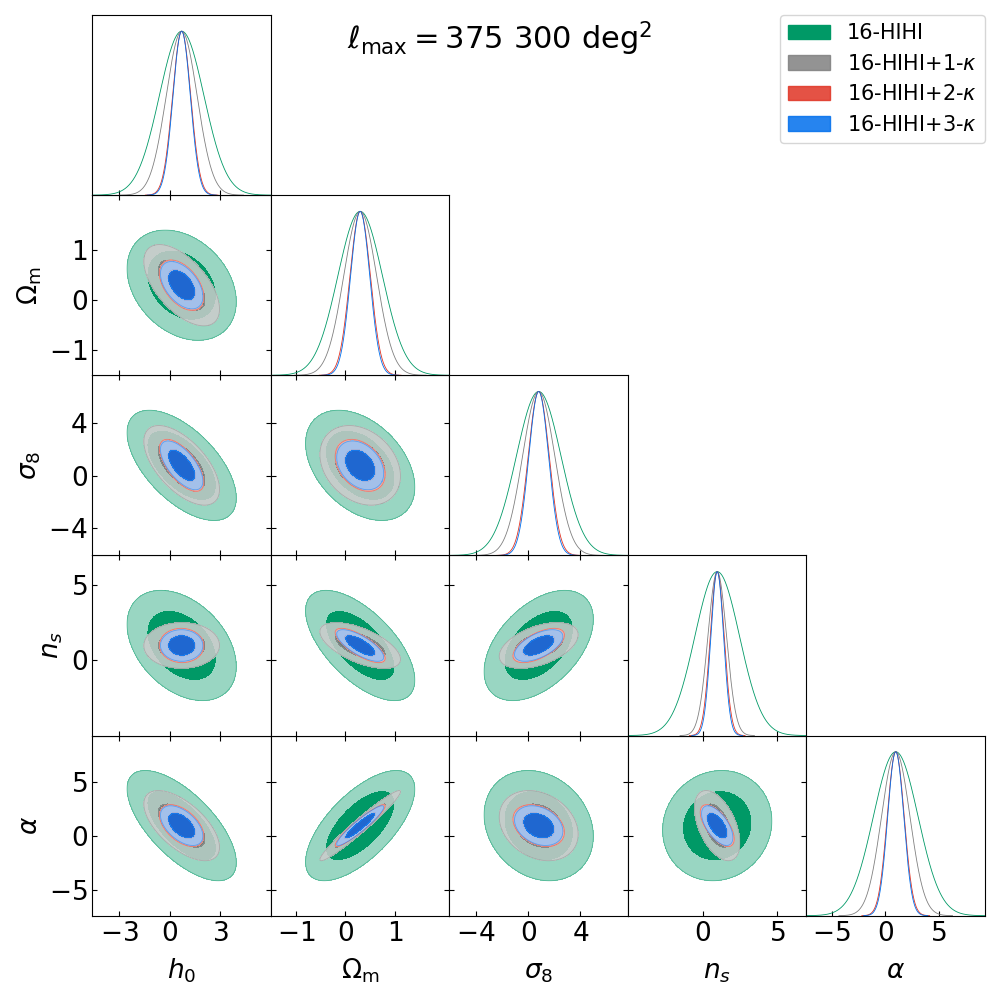}
\caption{Constraints on cosmological parameters and $b_\mathrm{HI}(z)$ for 2-point functions at 68\% and 95\% levels of confidence, for the small area 300 deg$^2$ case. This plot indicates that the feasibility of $\kappa$HI (pseudo) 2-point statistics in cosmological constraints is heavily affected by the statistical incompleteness due to having a small-sky survey.}\label{fig:bias1_cut}
\end{figure}

{Now let us consider the effect of sky coverage area on 2-point statistics of both HIHI auto and $\kappa$HI cross angular power spectra. We now consider a combined survey area 300 deg$^2$ of lensing and HI observations, comparable to current pathfinder intensity mapping surveys. As this area is much smaller than the full-sky case, we therefore now need to use the pseudo angular power ($\tilde{C}_\ell$) as an estimator of $C_\ell$.

Suppose the survey footprint of the observations can be expressed using the weight function $W(\hat{n})$. Normalised by the sky factor $f_\mathrm{sky}$ (the fraction of the sky covered by the data), the weight moments are given by:
\begin{equation}
f_\mathrm{sky}w_i = \frac{1}{4\pi}\int d\hat{n} W^i(\hat{n}),
\end{equation}
where $w_i$ represents the $i$-th moment of weighting. The power spectrum of the window function is:
\begin{equation}
\mathcal{W}_\ell = \frac{1}{2\ell + 1}\sum_m |w_{\ell m}|^2.
\end{equation}
For a spin-0 field $\zeta(\hat{n})$ weighted by $W(\hat{n})$, a spherical harmonic coefficient $\tilde{a}_{\ell m}$ can be expressed as~\citep{Kim:2010,Kim:2011}:
\begin{equation}
\tilde{a}_{\ell m} = \int d\hat{n}\zeta(\hat{n})W(\hat{n})Y^*_{\ell m}(\hat{n}) \approx \Omega_p \sum_p \zeta(p)W(p)Y^*_{\ell m}(p),
\end{equation}
where we approximate the integration over sky factor by the summation over pixel area with the surface density $\Omega_p$. The pseudo power spectrum estimator, $\tilde{C}_\ell$, is then:
\begin{equation}
\tilde{C}_\ell  = \frac{1}{2\ell+1}\sum_m |\tilde{a}_{\ell m}|^2.
\end{equation}
Similarly for spin-2 fields ($\gamma_1, \gamma_2$), we can obtain the coefficients, $\tilde{a}_{\pm2,\ell m}$, by:
\begin{equation}
\tilde{a}_{\pm2,\ell m} = \int[\tilde{\gamma}_1(\hat{n}) \pm i\tilde{\gamma}_2(\hat{n})]\,\,{_{\pm2}}Y^*_{\ell m}(\hat{n}) d\hat{n},
\end{equation}
where 
\begin{equation}
\tilde{\gamma}_1(\hat{n}) \pm i\tilde{\gamma}_2 = W(\hat{n})[\gamma_1(\hat{n}) \pm i\gamma_2(\hat{n})].
\end{equation}
Similarly to the full-sky case, $\tilde{a}^\mathrm{E}_{\ell m}$ and $\tilde{a}^\mathrm{B}_{\ell m}$ are then:
\begin{equation}
\tilde{a
}^\mathrm{E} = - (\tilde{a}_{2,\ell m} + \tilde{a}_{-2, \ell m})/2,
\end{equation}
\begin{equation}
\tilde{a}^\mathrm{B} = i(\tilde{a}_{2,\ell m} - \tilde{a}_{-2, \ell m})/2.
\end{equation}
The pseudo power spectra for E and B modes are:
\begin{equation}
\tilde{C}^\mathrm{E,B}_\ell = \frac{1}{2\ell+1}\sum_m |\tilde{a}^\mathrm{E,B}_{\ell m}|^2.
\end{equation}
The pseudo power spectra $\tilde{C}_\ell$ and true $C_\ell$ are related by the mode-mode coupling resulting from masking ($M_{\ell\ell'}$):
\begin{equation}
\langle \tilde{C}_\ell \rangle = \sum_\ell M_{\ell\ell'} \langle C_\ell \rangle.
\end{equation}
This kernel depends solely on the geometry of a cut-sky $\mathcal{W}_\ell$ and plays a crucial role in the pseudo-$C_\ell$ method. For details concerning the mode-mode coupling, see~\cite{Hivon:2002,Alonso:2019}.

We utilise NaMaster ~\citep{Alonso:2019}, which is a software package to calculate pseudo-$C_\ell$ for any spin fields, to evaluate HIHI and $\kappa$HI pseudo-$C_\ell$ for 300 deg$^2$ of our simulations above, within a masked region RA = [0, 30] deg and dec = +[0,10] deg, for the same redshift bins (see Table~\ref{tab:redshift_bin}). We choose the same $\ell$ bins as before up to $\ell<375$ with $\theta_\mathrm{beam} = 1$ deg convolution, and calculate the cut-sky covariance matrix using NaMaster. 
Fig.~\ref{fig:bias1_cut} shows the cosmological constraints feasible of this scenario. Comparing the results to the full-sky case (Fig.~\ref{fig:bias_1l300}), we can see that the statistical incompleteness due to having a cut-sky survey reduces the feasibility; with 300 deg$^2$ sky, we cannot detect the HIHI cosmic signal. However, incorporating the cross-correlation with weak lensing improves the significance of the joint statistics (Fig.~\ref{fig:bias1_cut}). }
\section{Instrument Noise}\label{sec:instrument_noise}

In this section we consider the instrument noise for both lensing and HI surveys for the current state of art case. We will consider the expected thermal noise for a single-dish survey for HI measurement. 
\subsection{Single Dish Thermal Noise}\label{subsec:single_dish_noise}
We begin by examining the noise on 3D measured power spectra for the HI auto-correlation ($P$). This discussion is based on the works of ~\citep{Battye:2013,Santos:2015,Bigot-Sazy:2015}. Subsequently, we derive the root mean square (rms) thermal noise expected on IM maps from the power spectra, which we will use to assess the effect of realistic noise on the ability to detect the lensing-HI cross-correlation.

The expected uncertainty ($\sigma_p$) on the power spectrum $P$ can be estimated by evaluating its expected second moment. By calculating the ratio between $\sigma_p$ and $P$  averaging  over the radial wavenumber bin size $\Delta{k}$, we can estimate the uncertainty in the IM map measurements. Following this procedure the error on $P$  can be estimated by the following expression~\citep{Seo:2010,Feldman:1994,Battye:2013}:
\begin{equation}
\frac{\sigma_p}{P} = 2 \sqrt{\left(\frac{2\pi}{3}\right) V_{\text{sur}} \frac{1}{4\pi k^2 \overline{\Delta k}}} \bigg(1 + \frac{\sigma^2_{\text{pix}} V_{\text{pix}}}{ \left[\overline{T}(z)\right]^2 W(k)^2 P} \bigg),
\end{equation}
where $V_\mathrm{sur}$ is the volume of survey, $W(k)$ is windows function, $\bar{T}(z)$ is the average temperature at given redshift $z$ and $V_\mathrm{pix}$ is the pixel volume.   In this work we shall ignore the contribution of shot noise and  only consider the contribution from pixel (thermal) noise $\sigma_\mathrm{pix}$. These parameters are discussed in detail in the following paragraphs.

It is essential to pick the proper size of $\Delta k$ to optimise the viability of a single dish. As we focus on the cosmic signal the acoustic scale should be  an aim. That means we require $\Delta k/k_A < 1$, where $k_A$ is the wavenumber of acoustic scale. The volume of the survey $V_\mathrm{sur}$ can be computed by
\begin{equation}
    V_\mathrm{sur} =\Omega_\mathrm{sur}\int^{z_\mathrm{max}}_{z_\mathrm{min}} dz \frac{dV}{dzd\Omega},
\end{equation}
where
\begin{equation}
\frac{dV}{dzd\Omega} = \frac{cr^2(z)}{H_0E(z)},
\end{equation}
where we assume a flat universe.
The window function $W(k)$ is set by the instrument's specification. As we map the $\delta_\mathrm{HI}$ from multiple redshift bins, we can ignore the contribution of radial directions in $W(k)$. However for the angular direction, this is not the case. Importantly, the angular resolution of the radio beam will define the noise level.  We can model $W(k)$ by:
\begin{equation}
    W(k) = \exp \bigg[ -\frac{1}{2}k^2r^2(z)\frac{\theta_\mathrm{FWHM}}{8\ln{(2)}}\bigg],
\end{equation}
where $\theta_\mathrm{FWHM}$ is the full width half max of angular resolution. Realistically,  $\theta_\mathrm{FWHM}$ depends on frequency ($\nu$). However, in this analysis we assume the variation of $\theta_\mathrm{FWHM}$ is small and negligible.  

The pixel volume, $V_\mathrm{pix}$, can be determined by
\begin{equation}
    V_\mathrm{pix} = \Omega_\mathrm{pix}\int^{z_\mathrm{c}+\Delta z}_{z_\mathrm{c}-\Delta{z}} dz\frac{dV}{dzd\Omega},
\end{equation}
where $z_\mathrm{c}$ is the central redshift and $\Delta{z}$ corresponds to  frequency width $(\Delta{\nu})$ and $\Omega_\mathrm{pix}$ is the pixel solid angle.

In radio astronomy we normally measure the signal in terms of power. The antenna temperature then generates the thermal noise; the pixel noise $\sigma_\mathrm{pix}$ can be approximated by~\citep{Santos:2015,Seo:2010},
\begin{equation}\label{eq:pix_noise_single}
    \sigma_\mathrm{pix} \approx \frac{T_\mathrm{sys}}{\varepsilon\sqrt{t_p2\Delta{\nu}}},
\end{equation}
where $t_p$ represents the observation time per pointing, $\varepsilon$ (approximately 1) signifies the efficiency of the telescope, meaning that almost no signal is lost when radio radiation is transmitted to the antenna. $T_\mathrm{sys}$ represents the system temperature, which includes
\begin{equation}\label{eq:T_sys}
    T_\mathrm{sys} = T_\mathrm{rx}+T_\mathrm{spl}+T_\mathrm{CMB}+T_\mathrm{gal},\end{equation}
where we ignore the contribution from the Earth's atmosphere. $T_\mathrm{spl}$ is the spill over from ground radiation (approximately 3K), $T_\mathrm{CMB}$ $\approx$ 2.73K and galactic temperature $T_\mathrm{gal}$ $\approx$ 25K (408 MHz/$\nu$)$^{2.75\mathrm{K}}$ ~\citep{Bacon:2018dui}.    The observing time per pointing $t_p$ relates to the total observation by $t_p$ = $t_\mathrm{tot} (\theta_\mathrm{B})^2/\Omega_\mathrm{sur}$, where $\theta_\mathrm{B}$ is an angular pixel size. 

As $\sigma_\mathrm{pix}$ is the rms thermal noise,  by definition its square is the power per pixel volume ($P_N/V_\mathrm{pix})$. Therefore, the 3D noise power spectrum ($P_N$) is then ~\citep{Battye:2013,Santos:2015}
\begin{equation}
    P_N = \sigma_\mathrm{pix}^2V_\mathrm{pix} = r^2y\frac{T^2_\mathrm{sys}\Omega_\mathrm{sur}}{2\varepsilon^2t_\mathrm{tot}},
\end{equation}
where
\begin{equation}
    y =  cH(z)^{-1}\frac{(1+z)^2}{\nu_{21}}.
\end{equation}
If we have $N_\mathrm{d}$ dishes where each dish has $N_\mathrm{b}$ beams, we can take less time for each pointing area. The noise power spectrum is then reduced to
\begin{equation}
    P_N(N_\mathrm{d},N_\mathrm{b}) = \frac{r^2yT^2_\mathrm{sys}\Omega_\mathrm{sur}}{2\varepsilon^2 t_\mathrm{tot}N_\mathrm{b}N_\mathrm{d}}.
\end{equation}
To determine the optimisation of a survey strategy, we can estimate the suitable $\theta_\mathrm{FWHM}$ and $\Omega_\mathrm{sur}$ that minimises $\delta{k}_\mathrm{A}/k_\mathrm{A}$ for acoustic scale $k_\mathrm{A}$.  This acoustic scale $k_\mathrm{A}$ can be estimated by following the work of  ~\cite{Blake:2003,Battye:2013},
\begin{equation}
    \frac{P(k)}{P_\mathrm{ref}} = 1+ A k\exp{\bigg[-\bigg(\frac{k}{0.1 h \mathrm{Mpc^{-1}}}\bigg)^{1.4} \bigg]}\sin{(2\pi k/k_\mathrm{A})},
\end{equation}
where $A$ is the overall amplitude which can be marginalised. The subscript 'ref' refers to reference cosmological parameters. 

We now calculate the thermal noise of a MeerKAT-like instrument $\sigma^\mathrm{MK}_\mathrm{pix}$ = $\sigma_\mathrm{T}$. In this case we consider a single dish telescope consisting of 64 dishes of 13.5m diameter, operating in UHF, L and S-band. The MeerKAT pilot survey by ~\cite{Wang:2021} focuses on L-band from 856 to 1712 MHz with 4096 frequency channels.  This pilot survey  has 10.5 hours observation time with approximately $\sim$ 200 deg$^2$ observation field~\citep{Wang:2021,Cunnington:2022a}. The summary statistics of this MeerKAT pilot survey are listed in Table~\ref{Table:MeerKAT_instru}. ~\cite{Wang:2021} show that for integrated frequency channels, they can achieve thermal noise $\sigma_\mathrm{T} \approx 2$ mK. 

\begin{table}
\centering
\caption{MeerKAT pilot survey specifications~\citep{Wang:2021}}\label{Table:MeerKAT_instru}
\begin{tabular}{cccc}
\hline
\\
$\Delta\nu$ & 0.2 MHz  \\\\
$N_{\Delta\nu}$& [200,250]  \\\\
 $T_\mathrm{rx}$& $7.5\times10^3 + 10^3(\nu[\mathrm{MHz}]/1000 - 0.75)^2$  [mK]\\\\
t$_\mathrm{tot}$ & 10.5 hours  \\\\
$z$ & [0.3885, 0.4623]  \\\\
$N_\mathrm{dish}$ & 64 \\\\
$T_\mathrm{sys}$ & 16 $\times 10^3$ mK\\\\
$N_\mathrm{pix}$ & 87500\\\\
$\theta_\mathrm{FHWM}$ & 1.48 deg\\\\
$\Omega_\mathrm{sur}$ & 200 deg$^2$
\end{tabular}
\end{table}

If we use Eq. (\ref{eq:pix_noise_single}) and (\ref{eq:T_sys}) together with Table~\ref{Table:MeerKAT_instru}, the expected $\sigma_\mathrm{pix}$ for a single frequency channel of MeerKAT pilot survey is
\begin{equation}\label{eq:sigma_mkpilot}
    \sigma_\mathrm{pix} (\Delta\nu = 0.2;10\rm{hr}) \approx 15\, \mathrm{mK},
\end{equation}
where we assume each dish has equal efficiency $\varepsilon$ =1 and consider only a single frequency channel $\Delta{\nu}$ = 0.2 MHz. If we consider the whole frequency range like~\citep{Wang:2021} the $\sigma_\mathrm{T}$ $\approx$ 2mK.

We now consider the case where the total observation time $t_\mathrm{tot}$ = 1000 hours and 250 frequency channels with $\Delta\nu$ = 0.2 MHz. 

\subsection{S/N of $\kappa\mathrm{HI}$}\label{subsec:KHI_rms}
In section~\ref{subsec:single_dish_noise} we have estimated the rms thermal noise for MeerKAT-like surveys similar to the current state-of-the-art.  In this section we explore the  future case, where the observation time $t_\mathrm{tot}$ can take longer than MeerKAT's pilot survey, and we assume a full-sky survey  to estimate the best possible S/N for weak lensing-intensity mapping ($\kappa$HI) 2-point statistics.  The estimate $\sigma^\mathrm{T}_\mathrm{pix}$ from Eq.~\ref{eq:sigma_mkpilot} is 15mK for one frequency channel, which is based on the specification of the current MeerKAT survey in Table~\ref{Table:MeerKAT_instru} and Eq.~\ref{eq:pix_noise_single}; to detect the cross-correlation we should find ways to reduce the $\sigma^\mathrm{T}_\mathrm{pix}$ as far as possible.

We first model the S/N of $\kappa$HI. We can consider the zero lag noise level for $\langle \kappa \Delta{T} \rangle$, i.e. where $\kappa$ and $\Delta T$ are measured in the same pixel. There is no reason why the  statistical noise  of $\kappa$ should be correlated with $\Delta{T}$. The noise for the cross-correlation will therefore be proportional to the product of $\sigma_\mathrm{T}$ and $\sigma^{n}_e$, where $\sigma_\mathrm{T}$ is HI thermal noise which can be estimated by Eq.~\ref{eq:sigma_mkpilot}. The rms noise for weak lensing $\sigma^n_e$ can be estimated by 
\begin{equation}\label{eq:sigma_ekl}
    \sigma^n_e =\frac{\sigma_e}{\sqrt{n_\mathrm{gal}}},
\end{equation}
where $\sigma_e$ is the variance of intrinsic galaxy ellipticities and $n_\mathrm{gal}$ is galaxy number per pixel.  For KiDS and DES-like surveys, $\sigma_e$ $\approx$ 0.3. The KiDS DR4 effective galaxy number density is $n_\mathrm{eff}$ = 0.325 arcmin$^{-2}$ for the whole redshift range~\citep{Heymans_kids:2021,Giblin_kids_lensing:2021}.  For a pixel size 0.25$^2$ deg$^2$, we find $\sigma^n_e$ $\approx$ 0.03.

The S/N of $\langle \kappa\mathrm{HI} \rangle$ then can be estimated by
\begin{equation}\label{eq:S2N}
    S/N = \frac{\sigma_\mathrm{HI}\sigma_\kappa}{\sigma_\mathrm{T}\sigma^n_e}\sqrt{N_\mathrm{pix}}.
\end{equation}
and we find that the rms lensing convergence signal $\sigma_\kappa$ is similar to the rms noise $\sigma^n_e$ on 0.25 deg scales~\citep{DES_shape_cat:2021,DES_Y3_cosmic_shear}. 

To estimate the signal of HI intensity mapping, we first recall that the HI brightness temperature fluctuations $\delta{T}_\mathrm{HI}$ can be expressed by:
\begin{equation}
    \delta{T}_\mathrm{HI} = \bar{T}_\mathrm{HI}(z)b_\mathrm{HI}(z)\delta_\mathrm{m}(\theta,z),
\end{equation}
where $\bar{T}_\mathrm{HI}(z)$ is an average temperature over angular position ($\theta$) for a given $z$ and $b_\mathrm{HI}(z)$ is HI bias.  As the power spectrum $P_\mathrm{HI}$ is the power of the temperature fluctuation $\delta T_\mathrm{HI}$,  the square root of $P_\mathrm{HI}$ per volume $V_\mathrm{sur}$ is effectively the root mean square of the HI true signal ($\sigma_\mathrm{HI}$),
\begin{equation}
    \sigma_\mathrm{HI} = \sqrt{P_\mathrm{HI}/V_\mathrm{sur}}.
\end{equation}

We consider a survey similar to the MeerKAT pilot survey  (0.39 < z < 0.46) over a moderately thick $\Delta{z}$ = 0.075 redshift bin, with 0.25$^2$ deg$^2$ pixel size, $V_\mathrm{sur}$ $\sim$ 4000 Mpc$^3$$h^{-3}$. Assuming the foregrounds and redshift space distortions have been appropriately dealt with, we can use
\begin{equation}
    P_\mathrm{HI} = \bar{T}^2_\mathrm{HI}b^2_\mathrm{HI}P_\mathrm{m}
\end{equation}
We assume an effective redshift of the survey $z_\mathrm{eff}$ = 0.42, $b_\mathrm{HI}$ =1  and $\bar{T}_\mathrm{HI}$ = 0.07 mK~\citep{Cunnington:2022a,Wang:2021,Santos:2015}.  In this case, the estimation of HI rms is then
\begin{equation}
    \sigma_\mathrm{HI}(k = 0.1, z = 0.42) = 5 \mu\mathrm{K},
\end{equation}
where we estimate at the $k$ = 0.1 $h^{-1}$Mpc scale. This estimation generally agrees with Table I in~\cite{Santos:2015} with slightly better signal rms, because~\cite{Santos:2015} uses smaller channel bins than the thick redshift bin we have here. 

$\sigma_\mathrm{T}$ is approximately 15 mK for a single frequency channel  given Table~\ref{Table:MeerKAT_instru}. If we stack over 200 $\Delta\nu$ channels and assume  10 hours observing time, then
\begin{equation}\label{eq:simgat200}
    \sigma_\mathrm{T} \approx 1.1 \, \mathrm{mK}.
\end{equation}
Then using equation (\ref{eq:S2N}), the estimation of S/N for $\kappa$HI 2-point statistics for  KiDS-like lensing surveys and MeerKAT for  $k$ = 0.1 $h^{-1}$ Mpc with a pixel size 0.25$^2$ deg$^2$ covering $\sim$ 200 deg$^2$ sky, is then
\begin{equation}
    S/N   \approx 0.24 .
\end{equation}
This means that by the current state of art, we expected to observe more instrument noise than cosmic signal from $\kappa$HI cross correlations at zero lag. By including cross-correlations at different angular separations, we expect a higher total signal-to-noise.

Note that this estimation is based on the pilot survey by MeerKAT which only contains 64 dishes of 13m diameter and only observes for 10hr; for the full operation of MeerKAT or SKA-Mid, we will have more dishes, longer observation time and more frequency channels.
If we assume a longer observation time such as  1000 hours (as is recommended by ~\citep{Zhang:2023}) and increase the number of frequency channels to 250, the estimation of $\sigma_\mathrm{MK}$ is then
\begin{equation}\label{eq:sigma_MK1000}
    \sigma^{1000}_\mathrm{MK} \approx 0.01 \, \mathrm{mK}.
\end{equation}

Now we have better S/N by one order of magnitude for a 3000 pixel sky. (If we use the pilot survey footprint ($\sim$ 200 deg$^2$), then S/N $\approx$ 2.4 for zero lag.)

We confirm this calculation by generating Gaussian random fields for   the instrument noise for both $\kappa$ and HI fields and add these noise maps using (Eq.~\ref{eq:sigma_mkpilot} and~\ref{eq:sigma_ekl}) to the simulations described in section~\ref{sec:fisher}.  We utilise the full sky maps with NSIDE=128. The pixel number for this resolution is $N_\mathrm{pix}$ = 196608. Hence the estimation of S/N for this configuration is then,
\begin{equation}\label{eq:S2N_1000_1}
    S/N^{1000}_\mathrm{full} \approx 22.
\end{equation}
However, when we take the LoS foreground subtraction into account, the signal would be reduced by a factor of $\sim$ 3. This means
\begin{equation}\label{eq:s2n_w_foreground}
    S/N^{1000}_\mathrm{full} \longrightarrow \simeq7. 
\end{equation}
This estimation indicates that we would expect to observe the true signal at zero lag for 1000 hour exposure and large sky surveys such as SKA.



\subsection{Fisher Analysis}\label{subsec:Fisher_w_noises}
We now consider the feasibility of $\kappa$HI and HIHI 2-point statistics for cosmological constraints in the presence of these current noise levels. In the S/N analysis of $\kappa$HI (sec.~\ref{subsec:KHI_rms}), we only considered the case where $\Delta\nu_\mathrm{total} = 250\times0.2$ MHz. However to compare results in sec.~\ref{subsec:bias_analysis}, we will adjust $\Delta{\nu}$ to match $z_\mathrm{HI}$ in Table.~\ref{tab:redshift_bin}.  

\begin{figure}
\centering
\includegraphics[width=0.5\textwidth]{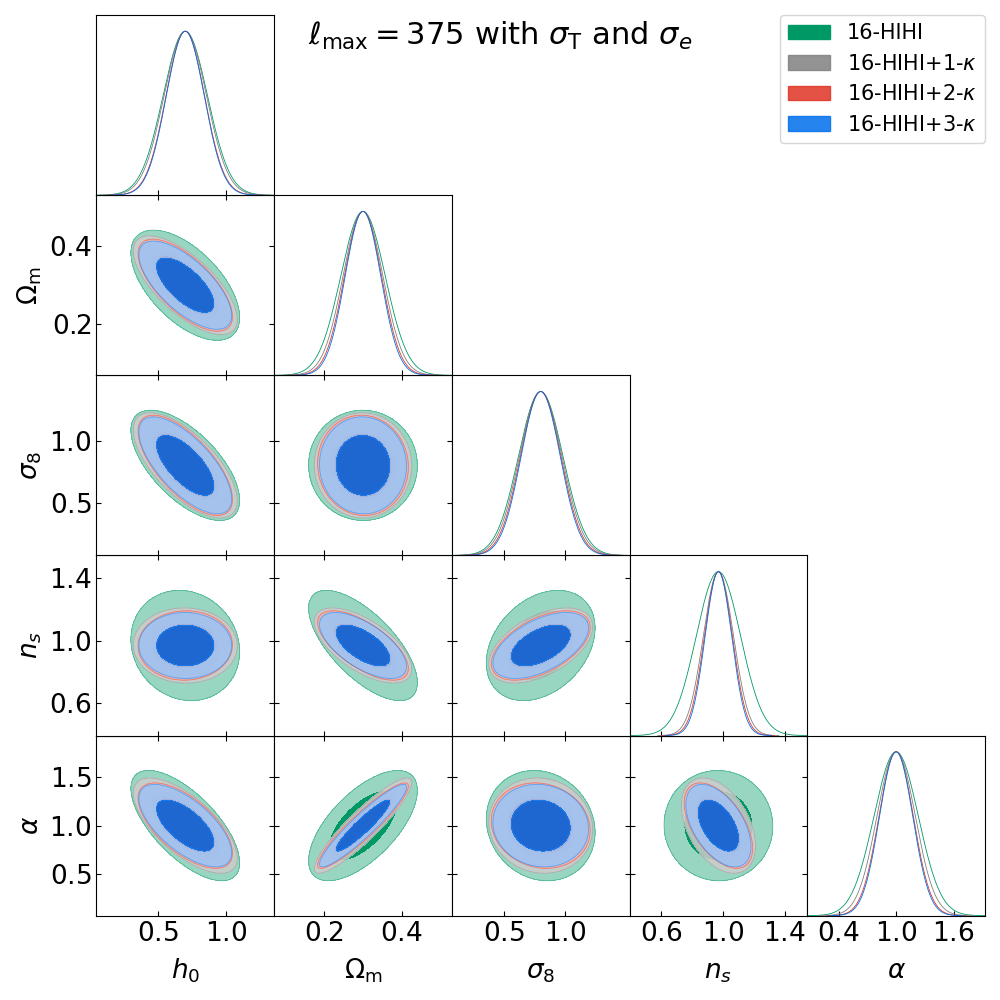}
\caption{The likelihood contours for our multi-bin analysis with HI bias model 1 with $\ell_\mathrm{max}=375$ for full-sky case including the instrument noise contributions; the contours show 68$\%$ and 95$\%$ confidence levels. Comparing to Fig.~\ref{fig:bias_1l300} where we ignore instrument noise contributions, we can see that $h_0$ is now poorly constrained.}\label{fig:bias_noise1l300}
\end{figure}

\begin{figure}
    \centering
    \includegraphics[width=0.5\textwidth]{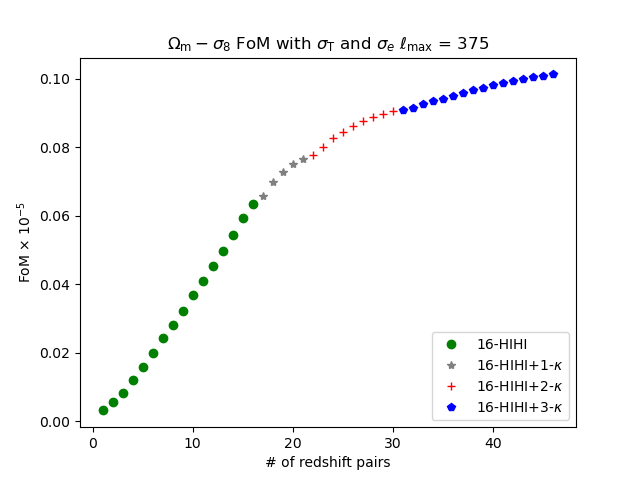}
    \caption{Figure of Merit for $\sigma_8 - \Omega_m$ constraints including instrument noise; the  horizontal axis is the number of redshift bin pairs for cosmological constraints. We show cumulative FoM when including increasing numbers of HI auto-correlation redshift bins (green); then increasing numbers of cross-correlations with convergence bins (grey, red, blue). Here $\ell_\mathrm{max} = 375$. Comparing this figure to Fig.~\ref{fig:FoMlmax300}, we find that the FoM is lower by 1 order of magnitude.}
    \label{fig:FoM_noise}
\end{figure}

We generate Gaussian white noise fields for both $\kappa$ and HI using the previous subsection's calculated amplitudes.  We note again that in this analysis, we consider only the full-sky case. The marginalisations of cosmological parameters are illustrated by Fig.~\ref{fig:bias_noise1l300}. Comparing this result to the no-noise case (see Fig.~\ref{fig:bias_1l300}) we can see that the cosmological feasibility of HIHI and $\kappa$HI are reduced significantly. We also show the resulting FoM in Fig.~\ref{fig:FoM_noise} and 2$\sigma$-constraints in Table~\ref{tab:cosmo_param_noise}. We note that the $\Omega_\mathrm{m}$ constraint shows a precise estimation, although it is degenerate with $\alpha$. Considering Fig.~\ref{fig:FoM_noise}, we can see that the maximum of FoM when considering instrument noise is one order of magnitude less than the no-noise case (see Fig.~\ref{fig:FoM}). Nevertheless, all FoM plots (see Fig.~\ref{fig:FoM},~\ref{fig:FoMlmax300}, and~\ref{fig:FoM_noise}) indicate that by combining $\kappa$HI likelihoods with HIHI, we can significantly enhance the feasibility of cosmological constraints from the HI intensity maps.

\begin{table}
	\centering
	\caption{Cosmological forecast from Fisher analysis including instrument noise contributions for 16-HIHI + 3-$\kappa$ correlations; the uncertainties on cosmological parameters and HI bias are quoted at 95$\%$ confident level. Here we consider only HI bias model 1. }
	\label{tab:cosmo_param_noise}
	\begin{tabular}{lccr} 
		\hline
		Parameters & HI bias model 1 \\
		\hline
		$\Delta{h_0}$  & $\pm$ 0.28\\
		$\Delta{\Omega_\mathrm{m}}$ & $\pm$ 0.09 \\
		$\Delta{\sigma_8}$ & $\pm$ 0.32 \\
		$\Delta{n_s}$ & $\pm$ 0.17  \\
        $\Delta\alpha$ & $\pm$ 0.34  \\
        \hline
	\end{tabular}
\end{table}
\section{Conclusions}
\label{sec:conclusions}
In this paper we have studied the 2-point statistics of lensing convergence and HI intensity mapping. 

We first presented the theoretical framework for calculating convergence-intensity mapping cross-correlations. Next, by using  realisations from an N-body simulation we have emulated HI intensity maps, and have shown that their cross-correlation with convergence maps from these simulations agree with our theoretical cross-correlation calculations.

We proceeded to study the effect of HI foreground removal on the 2-point functions. We model the effect of foreground removal by removing the mean along each line-of-sight, which effectively represents the largest radial mode. from our HI maps, following the method of~\citet{Cunnington:2019a}.
We then measure the post-removal cross-power; we find that the foreground removal modestly reduces the $\kappa$HI power spectrum signal, by a factor $A_\mathrm{clean}(z_\mathrm{HI}, z_\kappa, z_\mathrm{max})$.
In the case of forthcoming HI experiments that will measure HI at $z_\mathrm{max}$ < 1, $A_\mathrm{clean} (z_\mathrm{HI} < 0. 5, z_\kappa = 0.78, z_\mathrm{max}=1)$ is approximately 2.5 for our catalogues. 

In the following section, we utilised the Fisher matrix formalism to forecast best-case cosmological constraints for the convergence-HI probe, for the maximal case of full sky and subdominant telescope noise, but while including foreground removal. We calculated the Fisher matrix for $\kappa\kappa$, HIHI and $\kappa$HI 2-point functions by using the measured covariance matrices from Sec~\ref{sec:foreground}.

 We find that  a single redshift slice of the HI intensity map and $\kappa$ can constrain cosmological parameters for known bias (see Fig.~\ref{fig:3_set}), but when $b_\mathrm{HI}$ is a further parameter (or several), the few-slice $3\times2$ point functions do not sufficiently constrain the cosmological parameters compared to current cosmological surveys such as Planck and DES~\citep{Planck:2018:overview,Abbott:2018wog}. 

Hence, several cross-bin correlations are required in order for this probe to be of interest. In Sec~\ref{subsec:bias_analysis}, we have explored the use of several redshift bins for HI and convergence, together with the effect of $b_\mathrm{HI}$ on cosmological constraints. We consider both the current state of art where $\ell_\mathrm{max}< 400$, and the futuristic case where $\ell_\mathrm{max}$ > 1000.  Both cases show that a set of 2-point functions constrain the uncertainty in cosmological parameters to a comparable level with current experiments. All FoMs show that by including the cross-correlation of a lensing survey with the 21cm signal, we can improve the HI auto constraints.

We then examined the impact of a cut-sky survey. In this analysis, we evaluated the 2-point statistics based on a 300 deg$^2$ observed patch of sky. Due to the statistical incompleteness, detecting cosmic signals becomes marginal in this context.

In sec.~\ref{subsec:single_dish_noise}, we explore the instrument noise affecting lensing and HI galaxy surveys. The thermal noise of a single-dish survey was calculated. In this study, we focus on instruments similar to MeerKAT for radio observations and KIDs-like surveys for optical counterparts. Our analysis demonstrates the feasibility of detecting the $\kappa$HI cross-correlation, provided we have sufficient sky coverage and long exposure times for the radio measurements.

Even though we have shown  positive results for 2-point statistics between the $\kappa$ field and HI intensity map, there are important caveats that remain to be explored further:
\begin{itemize}
    \item In this study we have created HI intensity maps based on the assumption that they are linearly biased in relation to overdensity $\delta_\mathrm{m}$. A more realistic study should construct HI intensity maps by assigning HI mass $M_\mathrm{HI}$ to simulated halo catalogues. Also in future work the generated HI maps will be compared in detail to real data. 

    \item We have approximated the foreground cleaning by removing the mean fluctuation along the line of sight, which effectively represents the largest radial mode. More detailed emulation of the foreground cleaning will be studied in future.
    
    \item The cosmological and $b_\mathrm{HI}$ constraint predictions have been obtained by using the idealised Fisher matrix analysis; for real data,  Markov Chain Monte Carlo methods are required to deal with non-Gaussian likelihoods and realistic degeneracies between parameters.
        
    \item Only the $\Lambda$CDM model is considered in these parameter constraints. Further extensions (e.g. wCDM) should be considered in further work.
    
    \item Only 2-point statistics have been explored in this research. However the study by~\cite{Schmit:2019} shows that combining the bispectrum and power spectrum can  reduce the error of cosmological parameters by an order of magnitude compared to Planck.
    
    \item This study has only explored the low-redshift 2-point functions. The high redshift probes at the time of the Epoch of Reionization and the Cosmic Microwave Background are not taken into account. It is an interesting matter for future work to consider whether 2-point statistics between the EOR HI and CMB weak lensing can also be measured. 
\end{itemize}

In conclusion, $\kappa$-HI cross-correlations are an intriguing additional probe for cosmology, which are not destroyed by foreground removal. This probe will be available for measurement with forthcoming HI and lensing surveys this decade.

\section*{Acknowledgements}
We acknowledge studentship support from the Thai Royal Government for AS. We thank Tzu-Ching Chang (JPL), Yu-Wei Liao (ASIAA), Steve Cunnington (University of Manchester) and Utane Sawangwit (National Astronomical Research Institute of Thailand) for extremely helpful discussions. DB was supported by STFC Consolidator Grant ST/S000550/1. Calculations were made using Chanlawan High Performance Computer at the National Astronomical Research Institute of Thailand.

\section*{Data Availability}
In this work we use the N-Body simulations and weak lensing catalogues from~\cite{Takahashi:2017a}. The catalogues can be obtained at \url{http://cosmo.phys.hirosaki-u.ac.jp/takahasi/allsky\_raytracing}. The measured 2-point functions and Python codes for calculations in sections~\ref{sec:foreground} and~\ref{sec:fisher} are available in Github, at \url{https://github.com/AnutUoP/2020s-KappaHI}. Other data will be shared on reasonable requests to the corresponding author. 




\bibliographystyle{mnras}
\bibliography{Refs}








\bsp	
\label{lastpage}
\end{document}